\documentclass{aa}
\usepackage[varg]{txfonts}
\usepackage{natbib}
\usepackage{hyperref}
\bibpunct{(}{)}{;}{a}{}{,} 
\usepackage{graphicx}
\usepackage{xcolor}
\usepackage{booktabs}

\definecolor{myblue}{rgb}{0 0.349 0.702}

\hypersetup{
	colorlinks,
	linkcolor={myblue},
	citecolor={myblue},
	urlcolor={myblue}}

\title{AQUA: A Collection of H$_2$O Equations of State for Planetary Models\thanks{Tables \ref{Tab:eos_tab1}, \ref{Tab:eos_tab2} and \ref{Tab:eos_tab3} are only available in electronic form at the CDS via anonymous ftp to cdsarc.u-strasbg.fr (130.79.128.5) or via \url{http://cdsweb.u-strasbg.fr/cgi-bin/qcat?J/A+A/}}}
\titlerunning{AQUA: A Collection of H$_2$O Equations of State}

\author{Jonas Haldemann\inst{1}
	\and Yann Alibert\inst{1}
	\and Christoph Mordasini\inst{1}
	\and Willy Benz\inst{1}}

\offprints{Jonas Haldemann, \email{jonas.haldemann@space.unibe.ch}}

\institute{Department of Space Research \& Planetary Sciences, University of Bern, Gesellschaftsstrasse 6, 3012 Bern, Switzerland }

\date{Received 07 Mai 2020 / Accepted 18 September 2020}

\abstract {Water is one of the key chemical elements in planetary structure modelling. Due to its complex phase diagram, equations of state cover often only parts of the pressure -- temperature space needed in planetary modelling.}
{We construct an equation of state of H$_2$O spanning a very wide range from $0.1$ Pa to 400 TPa and 150 K to $10^{5}$ K, which can be used to model the interior of planets.}
{We combine equations of state valid in localised regions to form a continuous equation of state spanning over said pressure and temperature range.}
{We provide tabulated values for the most important thermodynamic quantities, i.e., density, adiabatic temperature gradient, entropy, internal energy and bulk speed of sound of water over this pressure and temperature range.  For better usability we also calculated density -- temperature and density -- internal energy grids.  We discuss further the impact of this equation of state on the mass radius relation of planets compared to other popular equation of states like ANEOS and QEOS.} {AQUA is a combination of existing equation of state useful for planetary models. We show that AQUA is in most regions a thermodynamic consistent description of water. At pressures above 10 GPa AQUA predicts systematic larger densities than ANEOS or QEOS. A feature which was already present in a previously proposed equation of state, which is the main underlying equation of this work. We show that the choice of the equation of state can have a large impact on the mass-radius relation, which highlights the importance of future developments in the field of equation of states and regarding experimental data of water at high pressures.} 


\begin{document}
\maketitle 
\section{Introduction}

Due to its abundance in the universe and its chemical properties, water is the key component in planetary models. It plays a major role during the formation and evolution of planets and at the same time it is thought to be a key ingredient in the emergence of life on Earth \citep{allen_origins_2003,wiggins_life_2008}. We find water not only in Earths hydrosphere but throughout the solar system, in the gas- \& ice-giants, their moons, in comets and other minor bodies \citep{grasset_water_2017}. Water is besides H/He also thought to be a dominant component in the atmospheres of giant exoplanets \citep{van_dishoeck_water_2014} while on smaller exoplanets it might form large oceans or thick ice-sheets \citep{sotin_massradius_2007}. The environments where water is expected to occur differ largely in their pressure and temperature conditions. To accurately model the interior structure of planets, a consistent description of the thermodynamic properties of water is needed over large pressure and temperature scales. Especially in anticipation of improved planetary radius measurements by space missions like CHEOPS \citep{benz_cheops_2017} or PLATO \citep{rauer_space_2018}, which require an accurate description of the planet major constituents, in order to constrain the planets bulk composition. In this work we combine multiple equations of state (EoS) all covering some part of the pressure and temperature (P -- T) space of water to construct an EoS useful for planetary structure modelling.

The phase diagram of water is highly diverse, including multiple ice phases and triple points. At ambient conditions the thermodynamic properties of water are well studied. The canonical reference, for pressures up to 1 GPa and temperatures of 1273 K, is provided by the International Association for the Properties of Water and Steam (IAPWS) in their IAPWS-R6-95 release \citep{wagner_iapws_2002} and for ice-1h in the IAPWS-R10-06 release \citep{feistel_new_2006}. Recently \citet{bollengier_thermodynamics_2019,brown_local_2018} expanded the validity region of their liquid-water EoS to higher pressures (2.3 GPa, respectively 100 GPa). At even higher pressures experimental data is more sparse and most work rely on \textit{ab initio} calculations to construct EoS. 

Recently \citet{mazevet_2019} published an \textit{ab initio} equation of state which spans over a pressure and temperature range useful for modelling the interior of giant planets. However due to the complications of low density \textit{ab initio} calculations the EoS is less accurate in low density regions ($\lesssim 1 $g/cm$^3$) of the P--T space\footnote{Although the results of \citet{wagner_iapws_2002} are used by \citet{mazevet_2019} to improve the provided fit towards lower temperatures.}. It also does not include any ice phases which are an important factor in models where the water is close or at the surface of planets. In this paper we show how one can combine the EoS of \citet{mazevet_2019} with other EoS which are more suitable at lower pressures and densities namely \citet{french_redmer_2015,journaux_holistic_2020,feistel_new_2006,wagner_iapws_2002,cea1_1994}. The resulting EoS can then be used to not only model water at high pressures and temperatures in planetary interiors as in \citet{mazevet_2019} but also to model planetary atmospheres, surfaces or moons, where water appears at lower temperatures and pressures.

We would like to note that we do not attempt to provide a better or novel EoS of water for the regions where the individual EoS were constructed for. The goal is rather to provide a continuous formulation of thermodynamic properties of water over large pressure and temperature scales. This is a crucial point in order to numerically solve the interior structure equations often used in planetary modelling.

This paper is structured in the following way. In section \ref{Sec:Methods} we describe the used EoS in more detail. We show where each EoS is used and how we transition between them. In section \ref{Sec:Results} we calculate the thermodynamic consistency of our approach and compare the resulting EoS with other EoS for water. In section \ref{Sec:MR} we use the AQUA-EoS to calculate mass radius relations for various boundary conditions and compare it against other common EoS. In the last section \ref{Sec:Conclusions} we discuss the major findings of this work. A public available version of the EoS in tabulated form (available as P-T, $\rho$-T and $\rho$-u grids.) can be found online at \url{https://github.com/mnijh/AQUA}.

\section{Methods}\label{Sec:Methods}
In the following section we describe how we combined the EoS of \citet{mazevet_2019} (hereafter M19-EoS) with other EoS which complement the M19-EoS at lower pressures and temperatures. All in perspective of developing a description of thermodynamic quantities used to model the interiors of planets and their satellites. The quantities we focus on are the density $\rho(T,P)$, the specific entropy $s(T,P)$, the specific internal energy $u(T,P)$, the bulk speed of sound 
\begin{equation}
w(T,P) = \sqrt{\left(\frac{\partial P}{\partial \rho}\right)_S}
\end{equation}
 and the adiabatic temperature gradient defined as
\begin{equation}
\nabla_\text{Ad} = \left(\frac{\partial \ln T}{\partial \ln P}\right)_S = \frac{\alpha_v T}{c_P\rho}\frac{P}{T}, \label{EQ:ad_grad}
\end{equation} where $\alpha_v$ is the volumetric thermal expansion coefficient and $c_P$ is the specific isobaric heat capacity. These quantities can be calculated from first or second order derivatives of a Gibbs or Helmholtz free energy potential. Finding a single functional form which accurately describes one of these energy potentials over the large phase space needed is very challenging and was not yet accomplished. Though there are many EoS describing the properties of H$_2$O in a localised region. We propose to use a selection of such local descriptions to construct an EoS of H$_2$O spanning from $0.1$ Pa to pressures of orders of TPa and temperatures between 100 K and \textbf{$10^5$} K. A similar method, though for a smaller P-T range and different EoS, was proposed by \citet{senft_impact_2008}. 

\subsection{Thermodynamic derivatives}
Each used EoS provides a functional form of either the Gibbs or Helmholtz free energy potential. Where the Gibbs free energy g(P,T) and Helmholtz free energy f($\rho$,T) are defined as
\begin{equation}
g(P,T) = u(P,T) + \frac{P}{\rho(P,T)} - T\cdot s(P,T)
\end{equation}
and
\begin{equation}
f(\rho,T) = u(\rho,T) - T\cdot s(\rho,T).
\end{equation}

As mentioned before, these potentials allow us to calculate all necessary thermodynamic properties by combination of first and second order derivatives of g(P,T) or f($\rho$,T) \citep{callen_thermodynamics_1985,thorade_partial_2013}. In case the EoS is formulated as a Gibbs potential g(P,T) we use the relations:
\begin{align}
	\rho(P,T) &= V(P,T)^{-1}=\left(\frac{\partial g(P,T)}{\partial P}\right)^{-1}_{T, N}, \label{EQ:gibbs_rho}\\
	s(P,T) &= -\left(\frac{\partial g(P,T)}{\partial T}\right)_{P, N},\label{EQ:gibbs_entro}\\
	u(P,T) &= g(P,T)+T \cdot s(P,T)-\frac{P}{\rho(P,T)}, \label{EQ:gibbs_energ}\\
	w(P,T) &= \sqrt{\frac{\left(\frac{\partial^2 g(P,T)}{\partial T^2}\right)_{P, N}}{\left(\frac{\partial^2 g(P,T)}{\partial T\partial P}\right)^2_{N} - \left(\frac{\partial^2 g(P,T)}{\partial T^2}\right)_{P, N} \left(\frac{\partial^2 g(P,T)}{\partial P^2}\right)_{T, N}}}, \label{EQ:gibbs_soundspeed}\\
	\alpha_v(P,T) &= \left(\frac{\partial^2 g(P,T)}{\partial T\partial P}\right)_{N} \rho(P,T) \label{Eq:gibbs_alpha}\\
	\intertext{and}
	c_P(P,T) &= -T\left(\frac{\partial^2 g(P,T)}{\partial T^2}\right)_{P, N}. \label{EQ:gibbs_cp}
\end{align}
to calculate the wanted quantities.
While in case of a Helmholtz free energy potential f($\rho$, T) we first solve which density corresponds to a given (P, T) tuple, using a bisection method and the relation
\begin{equation}
P(\rho,T) = \rho^2\left(\frac{\partial f(\rho,T)}{\partial \rho}\right)_{T, N}. \label{EQ:helm_p}
\end{equation}
Knowing the corresponding density $\rho(P,T)$ we then calculate the remaining quantities 
\begin{align}
	s(\rho,T) &= -\left(\frac{\partial f(\rho,T)}{\partial T}\right)_{\rho, N}, \label{EQ:helm_entro}\\
	u(\rho,T) &= f(\rho,T)+T\cdot s(\rho,T).
\end{align}
For $w$, $\alpha_v$ and $c_P$ we first calculate
\begin{align}
	K_T(\rho,T) &= 2\rho^2\left(\frac{\partial f(\rho,T)}{\partial \rho}\right)_{T, N} + \rho^3 \left(\frac{\partial^2 f(\rho,T)}{\partial \rho^2}\right)_{T, N}, \\
	\beta(\rho,T) &= \rho^2\left(\frac{\partial^2 f(\rho,T)}{\partial T \partial \rho}\right)_{N},
\end{align}
and then
\begin{align}
	w(\rho,T) &= \sqrt{\frac{K_T(\rho,T)}{\rho} - \frac{\beta(\rho,T)^2}{ \rho^2 \left(\frac{\partial^2 f(\rho,T)}{\partial T^2}\right)_{\rho, N}} }, \\
	\alpha_v(\rho,T) &= \frac{\beta(\rho,T)}{K_T(\rho,T)},\\
	c_P(\rho,T) &= -T\left(\frac{\partial^2 f(\rho,T)}{\partial T^2}\right)_{\rho, N} + T \frac{(\alpha_v(\rho,T)\,K_T(\rho,T))^2}{\rho K_T(\rho,T)}. \label{EQ:helm_cp}
\end{align}

\subsection{Transition between equation of states} \label{Sec:Combination}
When using multiple EoS for the same material, special care needs to be taken when and how to transition between EoS. Considering two EoS in P--T space, two major cases need to be distinguished. In the first case the two EoS describe two different phases of H$_2$O, hence a phase transition is expected to occur between the two EoS. The phase transition is, if present, the preferred location to transition between two EoS. By definition we expect discontinuities in the first and/or second order derivatives of the Gibbs free energy, hence no interpolation is needed to transition between two EoS. The location of the phase transition is either taken from experimental measurements or it is located where the two Gibbs energy potentials intersect \citet{poirier_introduction_2000}. The later approach is preferential in terms of consistency, though it might sometimes not recover the experimentally determined location of a phase transition.

\begin{figure}
	\resizebox{\hsize}{!}{\includegraphics{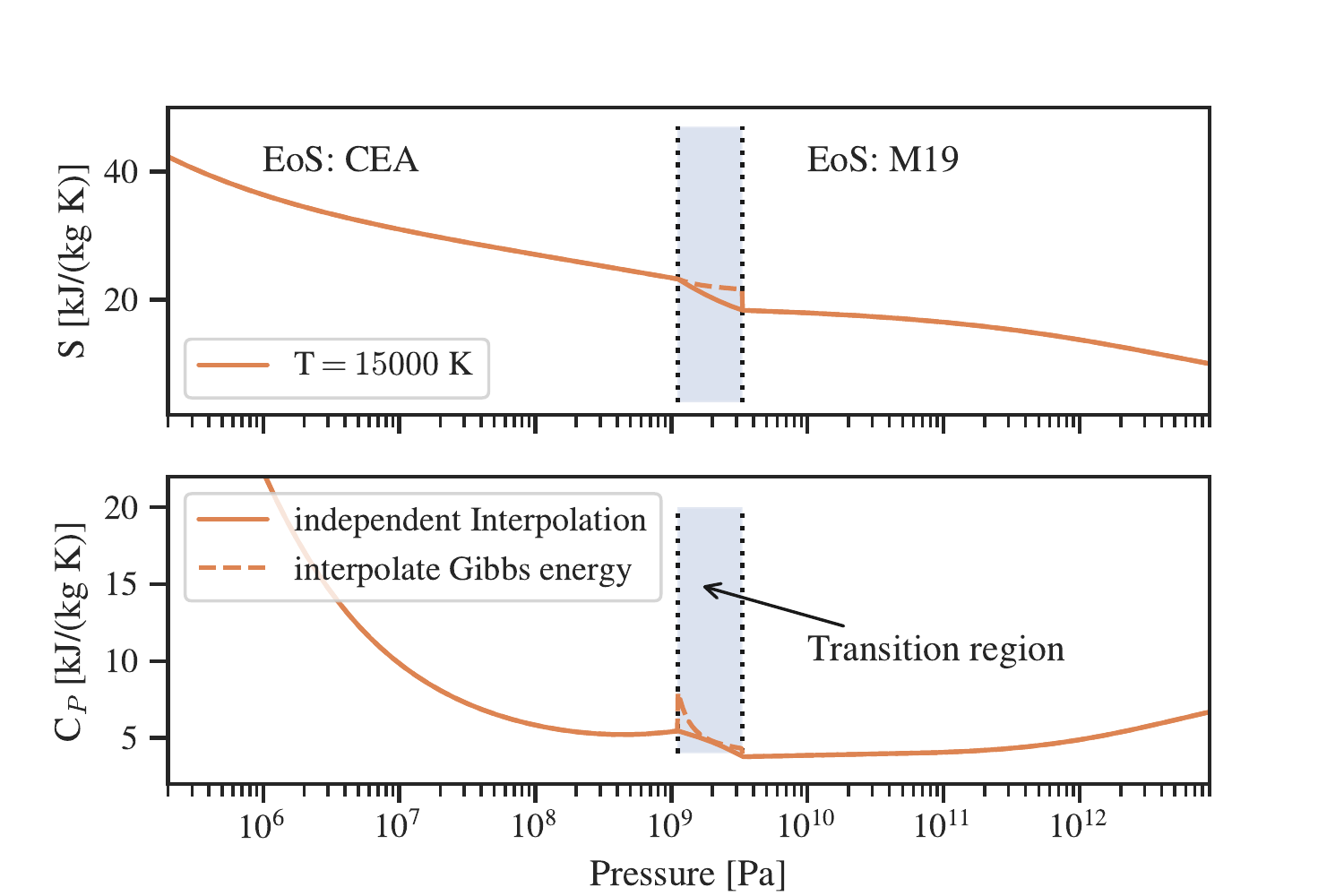}}
	\caption{Comparison between two possible methods of combining two equation of states. The solid lines correspond to interpolating all first and second order derivatives of the Gibbs potential independently. While for the dashed lines only the Gibbs free energy potential is interpolated and the derivatives are calculated thereof.}
	\label{Fig:Comp_method}
\end{figure}

In the second case, no phase transition is expected to occur between the two EoS. One could naively think that interpolating either the Gibbs or Helmholtz free energy potential of the two EoS and calculating all thermodynamic quantities from the interpolated potential would then be sufficient. But such interpolation can introduce new discontinuities in the first and second order derivatives of the respective energy potential (see Fig. \ref{Fig:Comp_method}). Even so when using special interpolation methods as proposed by, e.g., \citet{swesty_thermodynamically_1996} or \citet{baturin_interpolation_2019} which are thought to consistently evaluate tabulated EoS data, but are not thought for functionally different EoS. But besides interpolating the free energy potentials one can also interpolate all first and second order derivatives independently. Doing so, the aforementioned discontinuities are avoided, with the draw back that the thermodynamic consistency will not be guaranteed, i.e., the thermodynamic variables will show deviations from Eq. (3) - (18).
	
To illustrate this, we show in Fig. \ref{Fig:Comp_method} a transition between two EoS without interjacent phase transition. We see that when only the Gibbs free energy potential was interpolated (dashed lines), discontinuities were introduced in the entropy and specific heat capacity. While interpolating all first and second order derivatives (solid lines) results in a smooth behaviour. Hence a choice between a smooth but slightly thermodynamic inconsistent transition or a discontinuous but thermodynamic consistent transition has to be made. Assuming that both EoS are valid in their own region, we opt for the smooth transition, avoiding arbitrarily introduced discontinuities as shown in Fig. \ref{Fig:Comp_method}.
	
The two aforementioned cases lead to three methods which are used in this work to transition between EoS. The first method (in the following called Method 1) corresponds to the first case where a phase transition is expected between two EoS. There we locate the phase transition at the intersection of the Gibbs free energy potentials and change EoS at this location. If there is no interjacent phase transition (Method 2), then we define a transition region between the two EoS and interpolate in the first and second order derivatives of the Gibbs energy $g_X(P,T)$ using 
\begin{equation}
g_X = (1-\theta) \cdot g_{X}^\text{EoS$_1$} + \theta\cdot g_{X}^\text{EoS$_2$}. \label{EQ:interp1}
\end{equation}
The interpolation factor $\theta$ is calculated using either
\begin{equation}
\theta = (P-P_1)/(P_2-P_1),
\end{equation}
or
\begin{equation}
\theta = (T-T_1)/(T_2-T_1), \label{EQ:interp3}
\end{equation}
depending on the orientation of the transition region. The location and extend of the transition region is heuristically determined with the goal to reduce the introduced thermodynamic inconsistencies. In case the two neighbouring EoS were constructed such that they predict the same thermodynamic variables in an overlapping region (Method 3) then no special steps need to be taken when transitioning between the EoS. The two EoS can then be simply connected along a line within the overlap region.

\begin{table*}
	\caption{List of regions and the used EoS per region. As well as the method used to transition between two neighbouring regions (indicated as: region A $\leftrightarrow$ region B: Method X). The methods 1 - 3 are listed in §\ref{Sec:Combination}. }
	\label{Tab:eos_table}
	\centering
	\begin{tabular}{c l c c c}
		\hline\hline
		Region & Reference &Phase & \multicolumn{2}{c}{Transition Method} \\
		\hline
		1&\citet{feistel_new_2006} & ice-Ih & 1 $\leftrightarrow$ 2: Method 1 & 1 $\leftrightarrow$ 4: Method 1\\
		2&\citet{journaux_holistic_2020} & ice-II, -III, -V, -VI&2 $\leftrightarrow$ 5: Method 1 & 2 $\leftrightarrow$ 3: Method 1 \\
		3&\citet{french_redmer_2015} & ice-VII, -VII*, -X&3 $\leftrightarrow$ 7: Method 2&\\
		4&\citet{wagner_iapws_2002} & liquid \& gas \& supercritical fluid& 4 $\leftrightarrow$ 5: Method 3 & 4 $\leftrightarrow$ 6: Method 3\\
		5&\citet{brown_local_2018}&liquid \& supercritical fluid&5 $\leftrightarrow$ 3: Method 1 & 5 $\leftrightarrow$ 7: Method 2\\
		6&\citet{cea1_1994,cea2_1996}& gas & 6 $\leftrightarrow$ 7: Method 2 &\\
		7&\citet{mazevet_2019}&supercritical fluid \& superionic&&\\
		\hline
		
		\hline
	\end{tabular}
\end{table*}
\subsection{The pressure -- temperature regions}
\begin{figure}
	\resizebox{\hsize}{!}{\includegraphics{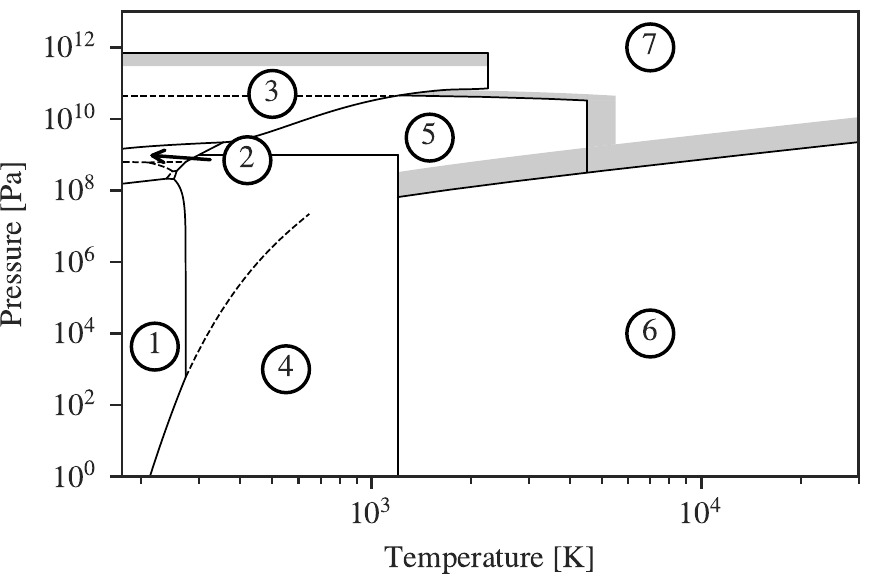}}
	\caption{Phase diagram of H$_2$O split into the seven regions listed in Table \ref{Tab:eos_table}. Most region boundaries (solid lines) follow phase transition curves. The dashed lines are phase transitions which are no region boundaries, i.e. the same EoS is used along the phase transition. The shaded areas show where neighbouring regions have to be interpolated. Region 7, i.e. where the EoS of \citet{mazevet_2019} is used, expands to temperatures up to $10^5$ K.}
	\label{Fig:WaterEOS_phases}
\end{figure}

We just saw that combining EoS can lead to potential inconsistencies. We therefore attempt to use as few EoS as possible. In a first step, the P--T space is split into seven regions for which a single EoS is chosen. The boundaries of the regions are located if possible along phase transition curves. 
An overview over the regions is given in Table \ref{Tab:eos_table}, where we also list which method was used to transition to the neighbouring EoS. Fig. \ref{Fig:WaterEOS_phases} further shows how the P--T space is split for the various EoS. The grey shaded areas in Fig. \ref{Fig:WaterEOS_phases} show where there is no physical phase transition between regions and interpolation (i.e., Method 2) is needed to assure a smooth transition of the thermodynamic variables.

\subsubsection{Region 1 (ice-Ih)}
The first region spans over the stability region of ice-Ih, bounded by the melting and sublimation curves as well as the ice-Ih/ice-II phase transition curve. 
For ice-Ih, the EoS from \citet{feistel_new_2006} is the canonical reference, adopted in the IAPWS-R10-06 release. It formulates a Gibbs energy potential and by design has a consistent transition when for the liquid and gas phase the EoS from the IAPWS-R6-95 release is used. The location of the melting and sublimation curves is then equal to the one described in \citet{wagner_2011}.

\subsubsection{Region 2 (ice-II, -III, -V and -VI)}
For region 2 we use the EoS described in the recent work of \citet{journaux_holistic_2020}. The EoS treats the ice-II, -III, -V and -VI phases and can consistently calculate the stability region of the various phases. To evaluate the EoS we use the \textit{seafreeze}-package\footnote{\url{https://github.com/Bjournaux/SeaFreeze}} of the same authors. \citet{journaux_holistic_2020} use local basis functions to fit a Gibbs energy potential to experimental data. The location of the phase transitions and region boundaries is calculated using the \textit{seafreeze}-package, which also uses a Gibbs minimisation scheme to locate the phase transitions. For the location of the ice-VI/ice-VII phase transition we use Method 1, i.e., calculating where the Gibbs potential of region 2 and 3 would intersect. We find that the following fit
 \begin{equation}
T_\text{67} = x_1 + x_2 \cdot (P/\text{Pa}) + x_3\cdot\log (P/\text{Pa}) + x_4\cdot\sqrt{(P/\text{Pa})}
\label{Eq:fit_67}
\end{equation}
parameterised the location of the phase transition between ice-VI and ice-VII up to the triple point at 2.216 GPa. The coefficients of Eq. (\ref{Eq:fit_67}) can be found in Table \ref{Tab:fit_melt7}. Reffering to the guidance of the \textit{seafreeze} package, it would be recommended to use the EoS of \citet{bollengier_thermodynamics_2019} for the liquid phase along with the ice phases of the \textit{seafreeze} package, in order to accurately recover the experimental location of the melting curves. 
But since the temperature range of \citet{bollengier_thermodynamics_2019} is restricted to 500 K we choose to use \citet{brown_local_2018} in the neighbouring region 5. We tested if using \citet{bollengier_thermodynamics_2019} would make a significant difference for the location of the melting curve. But changing to \citet{bollengier_thermodynamics_2019} for $T<500$ K only shifted the location given by Eq. (\ref{Eq:fit_melt7}) by a few Kelvin. Also the evaluated thermodynamic variables were equal, e.g., the maximal difference in density was 0.2\%). 

\subsubsection{Region 3 (ice-VII, ice-X)}
The third region is the stability region of the high pressure ice phases of ice-VII and ice-X, where we use the EoS by \citet{french_redmer_2015}. They provide a Helmholtz free energy potential which can be evaluated in the entire stability region of ice-VII and ice-X, up to 2250 K. The melting curve which separates region 3 towards region 5 and 7 was determined minimising the Gibbs free energy, i.e., Method 1. We found that the melting pressure can then be calculated using the following fit
\begin{equation}
\log_{10} P_\text{melt} = x_1\cdot(T/c)^{x_2}+x_3\cdot(T/c)^{-1}+x_4\cdot(T/c)^{-3} - 1. \label{Eq:fit_melt7}
\end{equation}	
Where $P_\text{melt}$ is in Pa and the coefficients of $x_i$ are given in Table \ref{Tab:fit_melt7}. The melting curve of ice-X starts at 1634.6 K and goes up to 2250 K, from where it follows an isotherm. This cut off at 2250 K is due to the limited range of the EoS, though it is similar to the experimental results of \citet{schwager_melting_2004}. 
\begin{table}
	\centering
	\caption{Coefficients for the fit of the melting  pressure of ice-VII and ice-X as well as the phase transition curve between ice-VI and ice-VII.}
	\label{Tab:fit_melt7}
	\begin{tabular}{c r r r}
		\hline
		\hline
		&melting ice-VII& melting ice-X& ice-VI/ice-VII\\
		\hline
		c & $355$ K & $1634.6$ K & \\
		$x_{1}$ & $2.6752$& $1.7818$ & $-1.4699\cdot 10^5$ K \\
		$x_{2}$ &  $-0.0269$ &  $0.2408$ & $6.10791\cdot 10^{-6}$ K\\
		$x_{3}$ & $-0.46234$ & $0.8310$ & $8.1529\cdot 10^3$ K\\
		$x_{4}$ &  $0.1237$ &  $-0.1444$ & $-8.8439\cdot 10^{-1}$ K\\
		\hline
	\end{tabular}
\end{table}
Between 700 GPa and 1.5 TPa ice-X is thought to undergo further structural changes until it transitions to the super-ionic phase \citep{militzer_2010}. 
Super-ionic water configurations are included in the \textit{ab initio} calculations of M19, though at higher temperatures than 2250 K. We tried adding also the EoS of super-ionic water as in \citet{french_equation_2009} to our description, but no good transition back to the M19-EoS was found. Therefore we decided that for pressures above 700 GPa we use the M19-EoS.

\subsubsection{Region 4 (gas, liquid and supercritical fluid)}
In region 4 we use the EoS from the IAPWS-R6-95 release \citep{wagner_iapws_2002}, the region spans over the entire liquid and the cold gas phase (<1200 K). The region boundaries follow the melting and sublimation curves from \citet{wagner_2011} until up to 1 GPa. The IAPWS-R6-95 does not cover H$_2$O vapour above 1273 K, we found that transitioning at 1200 K to the neighbouring region 6 results in a smooth transition (i.e., Method 3). The IAPWS-R6-95 is considered the canonical reference EoS for H$_2$O in this P--T region, it is formulated as a Helmholtz free energy potential and reproduces well experimental results. 

\subsubsection{Region 5 (liquid and supercritical Fluid)}
Since pressures above 1 GPa are outside of the validity region of the IAPWS-R6-95, we use the EoS by \citet{brown_local_2018}  for region 5. Through the usage of local basis functions to fit a Gibbs energy potential, \citet{brown_local_2018} provide an EoS which is appropriate for liquid and supercritical H$_2$O from 1 GPa to 100 GPa and up to $10^4$ K. \citet{brown_local_2018} used the IAPWS-R6-95 EoS, as a basis for their work, in order to transition between region 4 and 5 we simply switch the EoS at the boundary. The transitions to region 6 and 7 is discussed in their corresponding paragraph.

\subsubsection{Region 6 (ideal gas)}
At low densities H$_2$O vapour can be described as an ideal gas. Though thermal effects, like dissociation and thermal ionisation require a more complex treatment than pure ideal gas. In region 6 we use the CEA (Chemical Equilibrium with Applications) package \citep{cea1_1994,cea2_1996}, which can calculate the EoS of water at these conditions up to $2 \cdot 10^4$ K, including single ionisation and thermal dissociation. Besides the thermodynamic variables we calculate in this region also the mean molecular weight $\mu$, the dissociation fraction $x_\text{d}$ and the ionisation fraction $x_\text{ion}$, defined as
\begin{equation}
x_\text{d} = 1 - \frac{N_\text{H$_2$O}}{N} \label{EQ:x_disoc}
\end{equation}	
and
\begin{equation}
x_\text{ion} = \frac{N_e}{N}. \label{EQ:x_ion}
\end{equation}	
Where N is the total particle number, N$_\textbf{H$_2$O}$ the number of water molecules and N$_\textbf{e}$ the number of electrons.
The following species are considered when the CEA package is evaluated: H$_2$O, HO, H$_2$, H, O$_2$ and O, as well as the corresponding ions. In order to transition to region 5 and 7 we use Method 2 along the transition region shown in Fig. \ref{Fig:WaterEOS_phases}. 

\subsubsection{Region 7 (superionic phase and supercritical fluid)}
Region 7 corresponds to the M19-EoS. \citet{mazevet_2019} used Thomas-Fermi molecular dynamics (TFMD) simulations to construct a Helmholtz free energy potential up to densities of 100 g/cm$^3$, which corresponds to pressures of $\sim 400$ TPa. Although the TFMD calculations were performed up to temperatures of $5\cdot 10^4$ K we consider extrapolated values until $10^5$ K. This range in P and T should be sufficient to model most of the conditions in the interior of giant planets. 

Since there are no physical phase transitions between regions 4 to 7 we follow Method 2 in order to transition between the EoS. We tried to find transition regions in P--T space, where the difference between neighbouring EoS is minimal. 
The transition region between regions 5 and 7 is bound towards region 5 by
\begin{equation}
\log_{10} P_{5to7} = \log_{10} (42 \text{ GPa}) + \log_{10} (6 \text{ Pa})\frac{T/1000 K-2 }{18}\label{EQ:5to7}
\end{equation}
for temperatures between 1800 K and 4500 K, followed by an isothermal part until the boundary of region 6. While towards region 7 it is bound by $1.5\cdot P_{5to7}$ until 5500 K.  Similar for the border between region 6 and 7, for T >1000 K, the curves given by
\begin{equation}
\log_{10} P_{6to7} = \log_{10} (42 \text{ GPa}) + \log_{10} (6 \text{ Pa})\frac{T/1000 K-2 }{18}\label{EQ:6to7}
\end{equation}
and $3\cdot P_{6to7}$ bracket the transition region.  While the transition region towards region 3, is between 300 and 700 GPa up to 2250 K. The transition regions are indicated as grey areas in Fig. \ref{Fig:WaterEOS_phases}. We only evaluate the M19-EoS down to 300 K, hence at high pressures and T < 300 K, the M19-EoS will be evaluated at constant temperature. Though it is unlikely that water occurs at such conditions anyway.

\subsection{Energy and entropy shifts}
The Gibbs and Helmholtz free energy potentials, as well as the internal energy and the entropy are relative quantities. Hence they are defined in respect to a reference state. The IAPWS release for water provides two reference states the first arbitrarily sets the internal energy and entropy at the triple point (P$_t$ = 611.657 Pa, T$_t$ = 273.16 K) to zero, while the second recovers the true physical zero point (P$_0$ = 101325 Pa, T$_0$ = 0 K) entropy of ice-1h as calculated by \citet{nagle_lattice_1966}. M19 on the other hand, uses the second reference point from IAPWS, but sets the internal energy to be always positive.
In this work we will use the following reference state. At the zero point at P$_0$ = 101325 Pa, T$_0$ = 0 K, following \citet{nagle_lattice_1966}, the entropy is set to
\begin{equation} 
s(P_0,T_0) = 189.13 \, \text{J/(kg K)}.
\end{equation}
while the internal energy at the zero point is set to zero.
This means that for all used EoS, the entropy and energy values need to be shifted accordingly to ensure consistent transitions of the entropy and energy potentials.

Since the used reference state of \citet{french_redmer_2015} is not known, we shifted the energy potential in region 3 such that we recover the location of the ice-VI/VII/liquid triple point by \citet{wagner_2011}. See Table \ref{Tab:energy_shift} for an overview of the employed energy and entropy shifts.
\begin{table}[h]
	\caption{Overview over the energy and entropy shifts used to construct the AQUA EoS. }
	\label{Tab:energy_shift}
	\centering
	\begin{tabular}{c c c}
		\hline\hline
		Region & $\Delta s$ [J/(g K)] & $\Delta u$ [kJ/(g)] \\
		\hline
		1 &3.5164 & 0.632128736 \\
		2 &3.5164 & 0.632128736 \\
		3 &0.0    & 92.21378777 \\
		4 &3.5164 & 0.632128736 \\
		5 &3.5164 & 0.632128736 \\
		6 &0.0    & 16.59895404 \\
		7 &-0.5   &-2.467871264 \\
		\hline
	\end{tabular}
\end{table}

\section{Results}
\label{Sec:Results}
In the following section we discuss the properties of the  AQUA-EoS constructed with the method outlined in the last section. We validate said method and compare the thermodynamic variables calculated with the AQUA-EoS to other EoS.
 
\subsection{Tabulated equation of state}
For better usability we provide tabulated values of the AQUA-EoS on P--T, $\rho$--T and $\rho$--U grids. Since the regions and their boundaries are given in P--T, the $\rho$--T and $\rho$--u  grids are derived from the P--T grid. The fundamental P--T grid is calculated in the following way. For every point on the grid:
\begin{itemize}
	\item[(1)] Evaluate which region corresponds to the P,T values.
	\item[(2)] Calculate either the Gibbs or Helmholtz free energy given the regions EoS.
	\item[(3)] Evaluate either Eq. (\ref{EQ:gibbs_rho}) - (\ref{EQ:gibbs_cp}) or Eq. (\ref{EQ:helm_p}) - (\ref{EQ:helm_cp}) to calculate $\rho$, s, u, w, $\nabla_\text{Ad}$. 
	\item[(4)]If the P,T values are in region 6, calculate the ionisation and dissociation fractions Eq. (\ref{EQ:x_ion}), (\ref{EQ:x_disoc}) and the corresponding mean molecular weight.
	\item[(5)] If the P,T values are in a transition region, repeat steps (2) to (4) for the neighbouring region and transition between the two sets of thermodynamic variables as outlined in §\ref{Sec:Combination}
\end{itemize}

The tabulated AQUA-EoS is shown in Tables \ref{Tab:eos_tab1} - \ref{Tab:eos_tab3}. For the P--T table, we logarithmically sampled 70 points per decade from 0.1 Pa to 400 TPa and 100 points per decade from $10^2$ K to $10^5$ K. The rho-T table shares the same spacing along the temperature axis as the P--T table, while $\rho$ was sampled logarithmically with 100 points per decade from $10^{-10}$ kg/m$^3$ to $10^5$ kg/m$^3$. Similarly the rho--U table shares the same $\rho$ spacing as the $\rho$-T table, while the internal energy is logarithmically sampled with 100 points per decade from $10^5$ J/kg to $4\cdot10^9$ J/kg.  Due to its size, the tables are published in its entirety only in electronic form
at the CDS via anonymous ftp to cdsarc.u-strasbg.fr (130.79.128.5) or via \url{http://cdsweb.u-strasbg.fr/cgi-bin/qcat?J/A+A/}. The tables are also available to download under the link \url{https://github.com/mnijh/AQUA}.  

\subsection{Validation}
In order to validate the method of combining the selected EoS. We check for the thermodynamic consistency of the created tabulated EoS using the relation
\begin{equation}
\Delta_{Th.c.} \equiv 1-\frac{\rho(T,P)^2\left(\frac{\partial S(T,P)}{\partial P}\right)_T}{\left(\frac{\partial \rho(T,P)}{\partial T}\right)_P}\label{Eq:th_cons_text}
\end{equation} 
Which is a measure of how well the caloric and mechanical part of the EoS fulfil the fundamental thermodynamic relations used to derive Eq. (\ref{EQ:gibbs_rho})-(\ref{EQ:helm_cp}).
A similar approach was chosen, e.g., in \citet{timmes_accuracy_1999} and \citet{becker_ab_2014}. Though since we use P and T as natural variables for our EoS, the equation for the thermodynamic consistency measure differs from said authors which use $\rho$ and T. We derived Eq. (\ref{Eq:th_cons_text}) from the first law of thermodynamics in Appendix \ref{Sec:app_th_cons}. 
In Fig. \ref{Fig:WaterEOS_cons} we show Eq. (\ref{Eq:th_cons_text}) evaluated over the P--T domain. As one can expect from EoS based on Gibbs or Helmholtz free energy potentials, within the different regions our method preserves thermodynamic consistency. Some inconsistencies can be seen at  phase transitions between the ice phases as well as in the low pressure region of ice-1h, but they are rather small. The main inconsistencies are located between regions 5, 6 and 7. We would like to remind again that we attempt to create a formulation useful over large pressure and temperature scales. If an EoS is needed which is only used in a localised P--T domain, then other EoS will be more suitable. 
As already noted we evaluate the M19-EoS above 400 GPa and below 300 K at constant temperature, therefore in this region thermodynamic consistency is also not given, but again this region is unlikely to be encountered in planets.
Overall the method seems to deliver consistent results for the intended purpose of planetary structure modelling over a wide range of pressure and temperature. 
\begin{figure}
	\resizebox{\hsize}{!}{\includegraphics{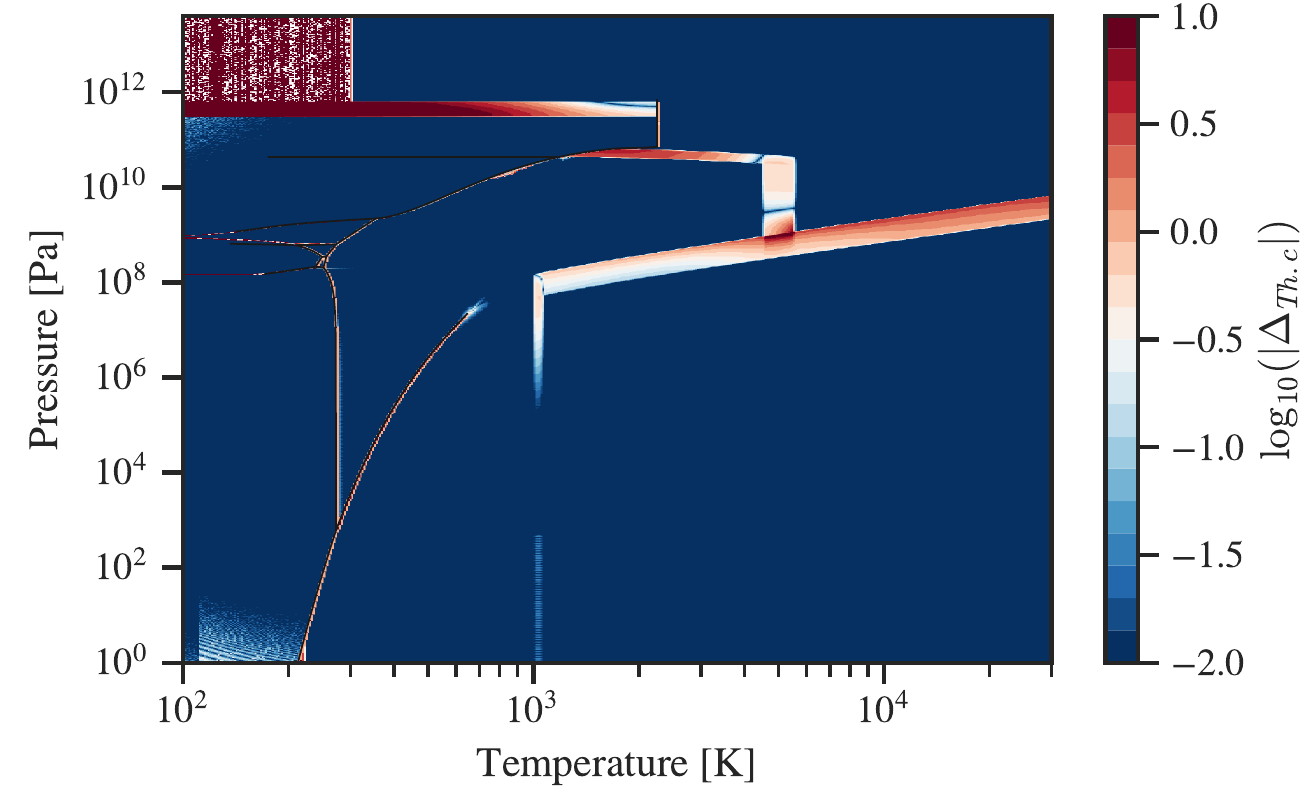}}
	\caption{Thermodynamic consistency measure $\delta_{Th.c.}$ defined in Eq. (\ref{Eq:th_cons_text}), as a function of pressure and temperature. Along phase transitions, the region boundaries and around the critical point deviations from the ideal thermodynamic behaviour can be seen. 
	The rectangular patch in the top left originates from evaluating the M19-EoS at constant temperature.}
	\label{Fig:WaterEOS_cons}
\end{figure}
\subsection{Density $\rho(P,T)$}
\begin{figure}
	\resizebox{\hsize}{!}{\includegraphics{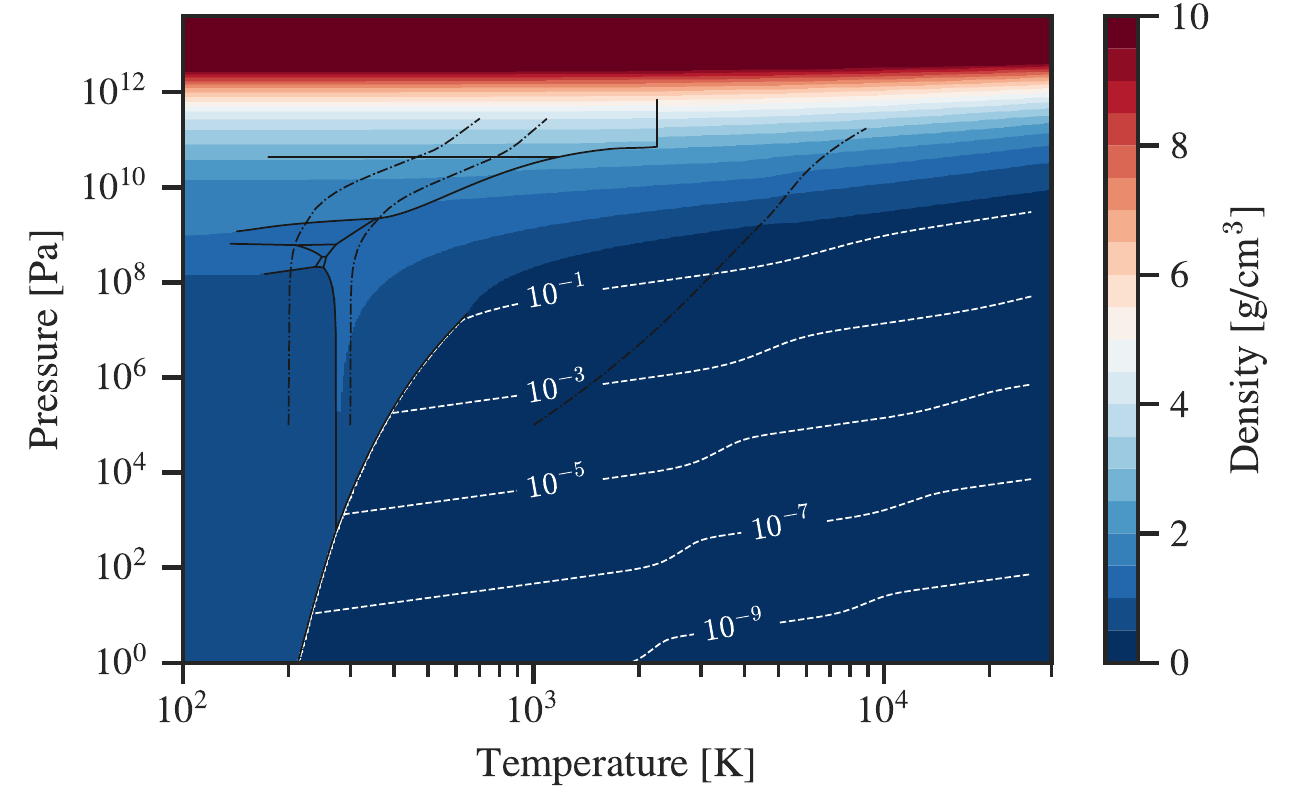}}
	\caption{Density of H$_2$O as a function of pressure and temperature calculated with the collection of H$_2$O EoS of this work. The various EoS used to generate this plot are listed in Table \ref{Tab:eos_table}. The solid black lines mark the phase transition between solid, liquid and gaseous phase. The white dashed lines are the density contours for the region where the density is below unity. The dot dashed black lines are adiabats calculated for a $5M_\oplus$ sphere of pure H$_2$O for different surface temperatures of 200 K, 300 K and 1000 K. }
	\label{Fig:WaterEOS_rho}
\end{figure}

From all studied thermodynamic variables, the $\rho$ -- P -- T relation of H$_2$O will have the biggest impact on the mass radius relation of planets. 
In Fig. \ref{Fig:WaterEOS_rho} we plotted $\rho$(P, T) using the AQUA-EoS from 1 Pa to 400 TPa and 150 K to \textbf{$3\cdot 10^4$} K. At higher temperatures, anyway only the M19-EoS contributes to the AQUA-EoS, so we forwent to expand the plot to this P--T region. The solid lines show the phase boundaries while the dashed lines are the density contours. Overlaid are with dot-dashed lines the adiabatic P,T profiles of a $5M_\oplus$ sphere of pure water for different surface temperatures of 200 K, 300 K and 1000 K.
\begin{figure}
	\resizebox{\hsize}{!}{\includegraphics{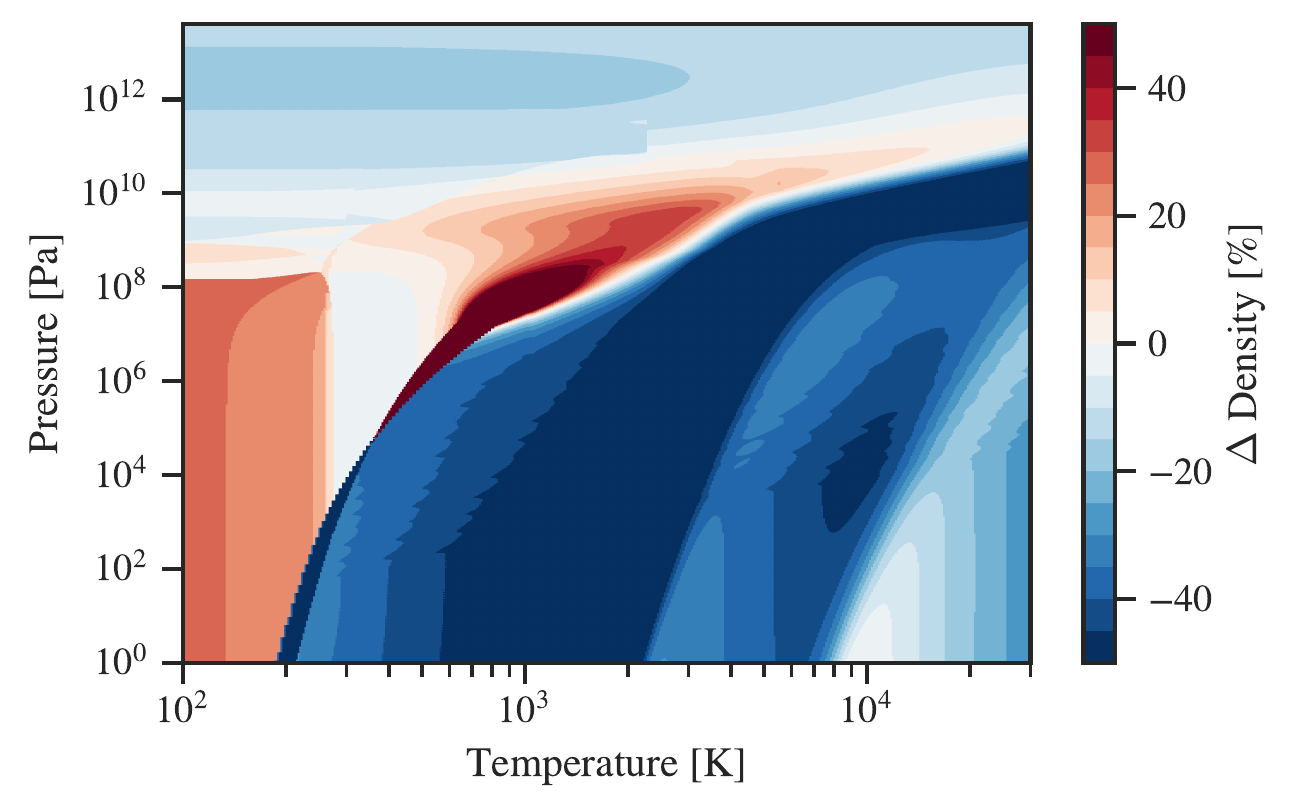}}
	\resizebox{\hsize}{!}{\includegraphics{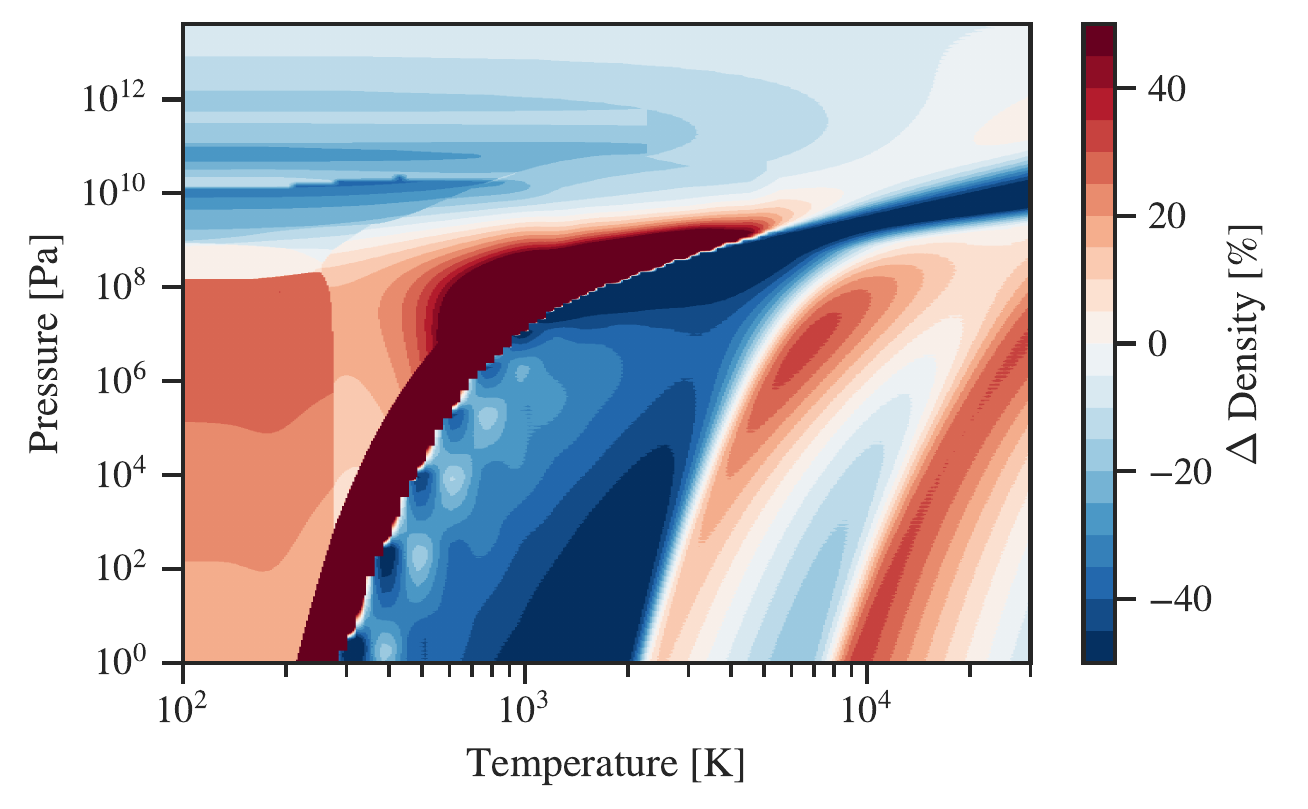}}
	\resizebox{\hsize}{!}{\includegraphics{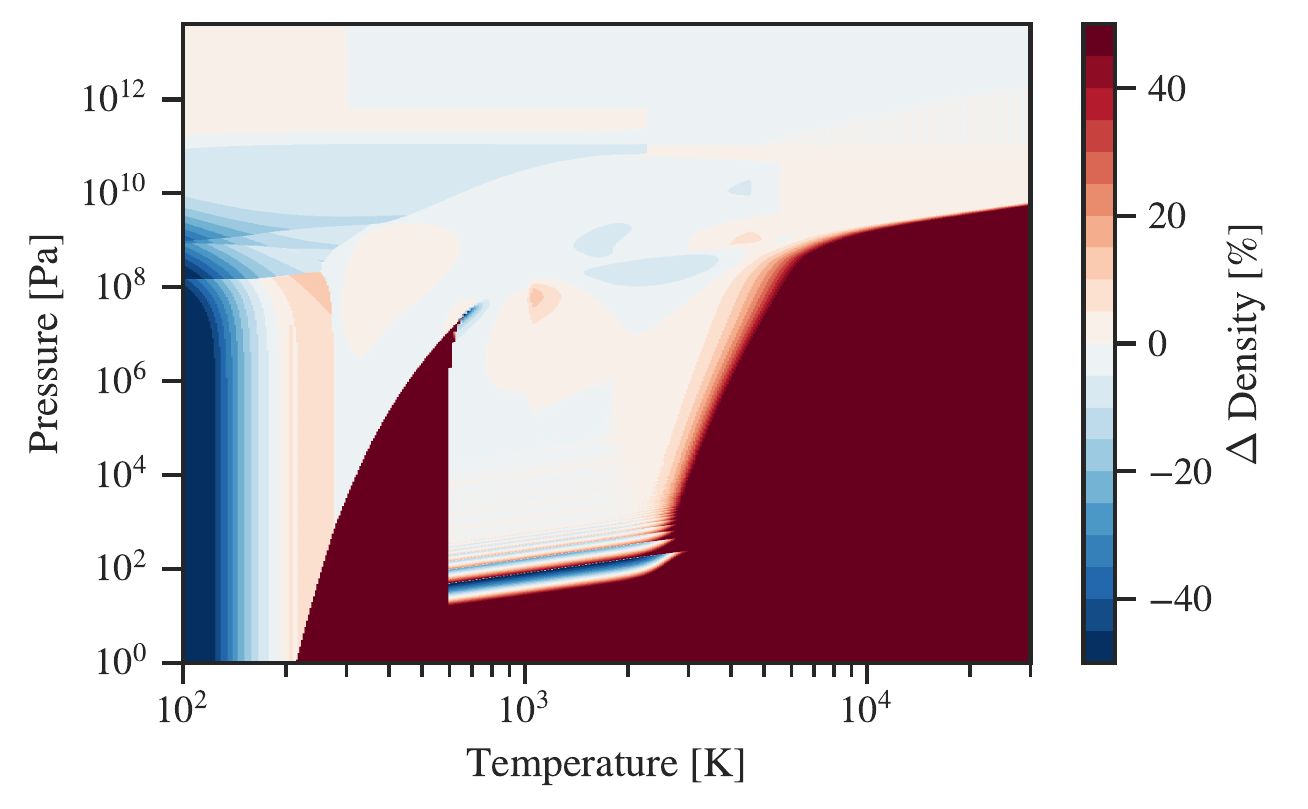}}
	\caption{Difference in density between the AQUA-EoS and ANEOS (top panel), QEOS (middle panel) and the M19-EoS (bottom panel). A positive difference means that the AQUA-EoS predicts a lower density in the specific location compared to the corresponding EoS of each panel.}
	\label{Fig:diff_rho}
\end{figure} 

In Fig. \ref{Fig:diff_rho} we compare the AQUA-EoS against common EoS for water used in planetary science, i.e. ANEOS (by \cite{thompson_aneos_1990} using parameters for water as in \citet{mordasini2020}), an improved version of QEOS (priv. comm. and \citet{vazan_helled_2013,vazan_explaining_2020}) and against the pure M19-EoS evaluated also at lower pressures and temperatures. Each panel in Fig. \ref{Fig:diff_rho} shows the density difference in percent between the AQUA-EoS and the corresponding EoS of each panel calculated using 
\begin{equation}
\Delta \rho(T,P)) = 100\cdot\frac{\rho_i(T,P)-\rho_\text{AQUA}(T,P)}{\rho_\text{AQUA}(T,P)}.
\label{EQ:diff}
\end{equation}
Where the index $i$ represents the EoS against which the difference in density is calculated. We note that both ANEOS and QEOS predict consistently lower densities at high pressures (>10 GPa). In contrary the density of ice-Ih is predicted higher than in the AQUA-EoS or as in \citet{feistel_new_2006}. In the gas phase below 1 GPa ANEOS predicts continuously lower densities, similar differences are seen for QEOS except above 2000 K where QEOS predicts slightly larger densities. QEOS has a significant shift in the location of the vapour curve and the location of the critical point.

For comparison we also evaluated the M19-EoS over the same P--T range. As expected no difference is seen at high pressures, since the same EoS is evaluated. At low pressures the density difference in the ice phases is visible and also the large differences in the gaseous low density regions below 1 GPa.

\subsection{Adiabatic temperature gradient $\nabla_\text{Ad}(P,T)$}
\begin{figure}
	\resizebox{\hsize}{!}{\includegraphics{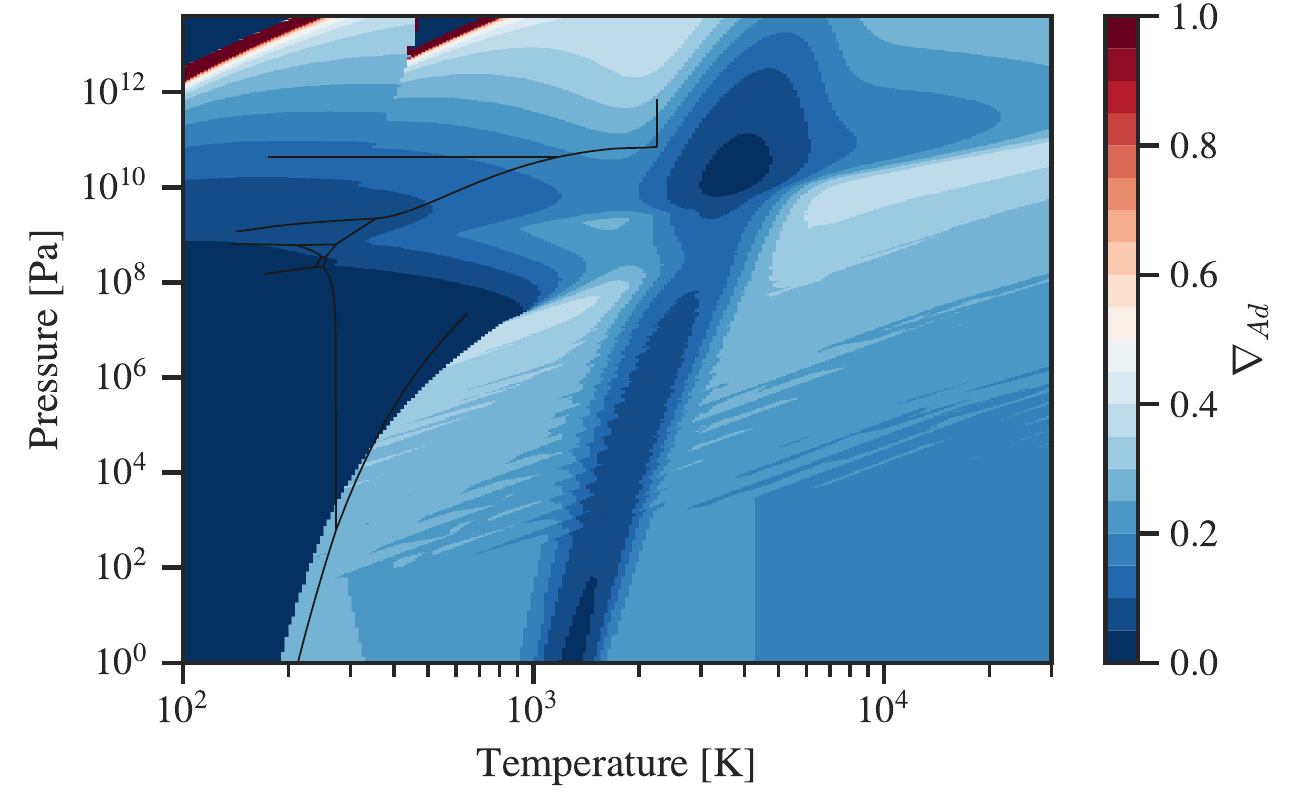}}
	\resizebox{\hsize}{!}{\includegraphics{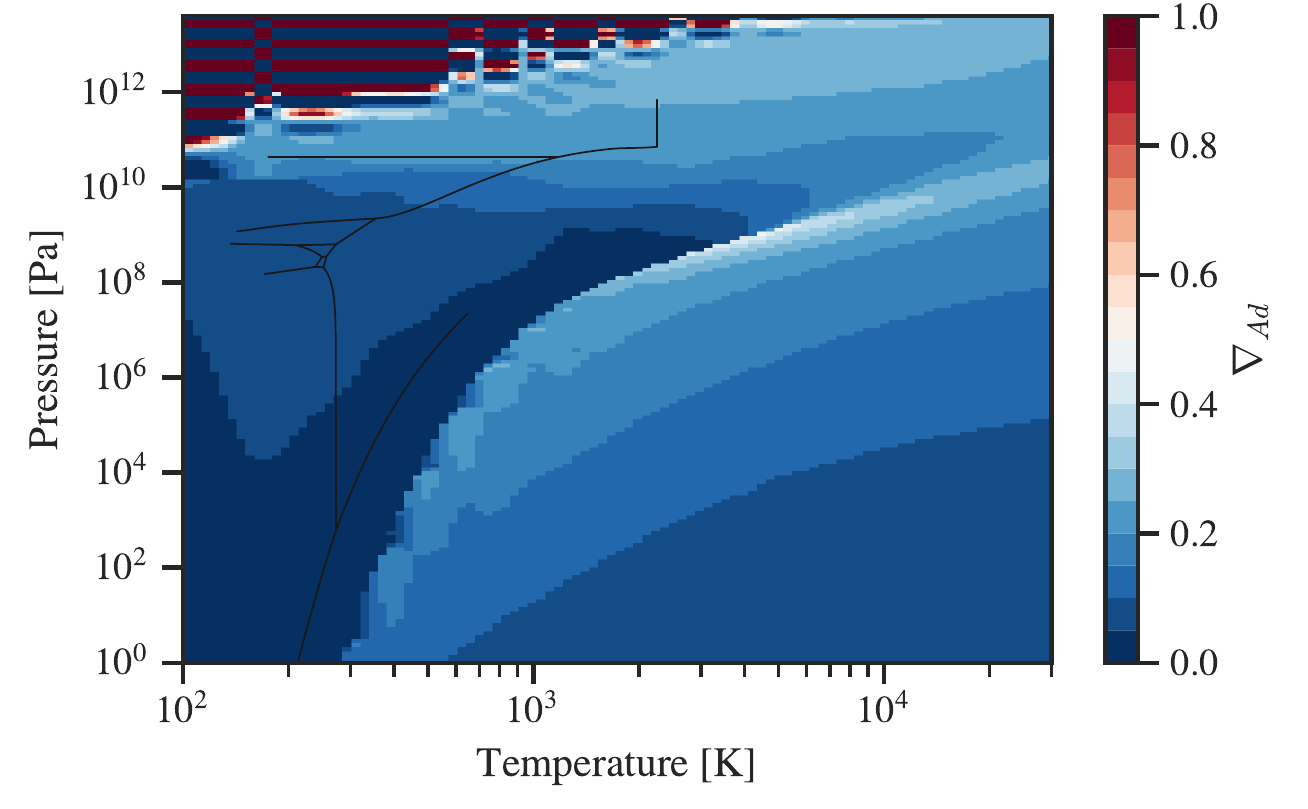}}
	\resizebox{\hsize}{!}{\includegraphics{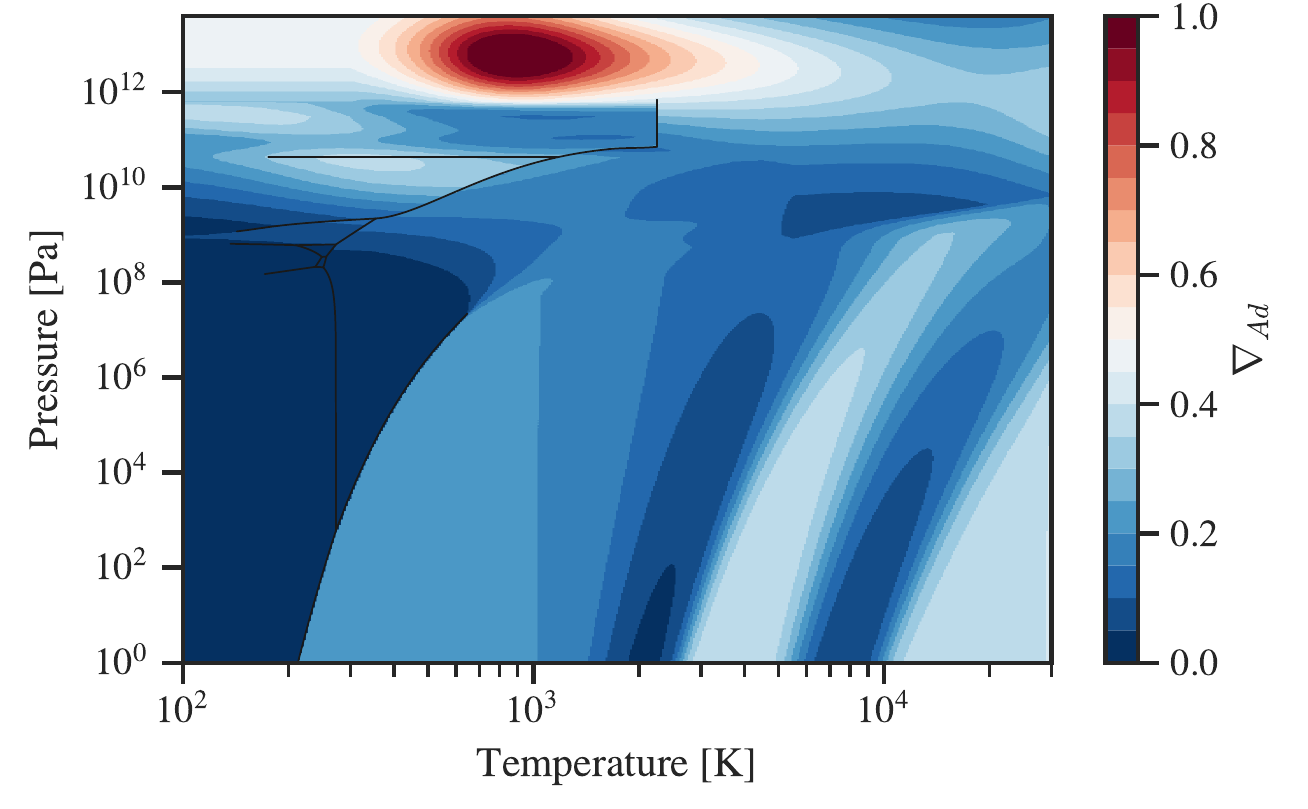}}
	\caption{Adiabatic temperature gradient of ANEOS (top panel), QEOS (bottom panel) and the AQUA-EoS (bottom panel) as a function of pressure and temperature. The black lines are the phase boundaries as in Fig. \ref{Fig:WaterEOS_rho}.}
	\label{Fig:WaterEOS_ad_grad}
\end{figure}

The dimensionless adiabatic temperature gradient $\nabla_\text{Ad}(P,T)$ is a key quantity to study the convective heat transport of a planet. In Fig. \ref{Fig:WaterEOS_ad_grad} we show the $\nabla_\text{Ad}(P,T)$ of the AQUA-EoS, ANEOS and QEOS.  
Compared to the the AQUA-EoS, the adiabatic gradient of ANEOS in the ice-Ih, liquid and cold gas phase is similar, while QEOS shows a larger gradient and a shifted vapour curve. In the gas phase, ANEOS shows a region of low adiabatic gradient between 1000K and 1100K. While for AQUA-EoS a similar feature caused by the thermal dissociation is visible but at higher temperatures. At the same time ANEOS does not include thermal ionisation effects which cause the second depression of $\nabla_\text{Ad}$ in AQUA between 6000 K and 10000 K. In QEOS none of these features are present. Since in both AQUA-EoS and ANEOS, the liquid and low pressure ice regions $\nabla_\text{Ad}(P,T)$ is close to zero. This will lead to an almost isothermal temperature profile. Therefore any adiabatic temperature profile starting in one of these regions will stay in its solid or liquid state until it eventually reaches the high pressure ices phases. Starting in the vapour phase will cause the temperature profile to be steep enough to remain in the vapour phase and then transition to the supercritical region. All EoS show numerical artefacts at low temperatures and high pressure, though it is unlikely that planetary models will need this part of P--T space.

\subsection{Entropy $s(P,T)$, internal energy $u(P,T)$}
\begin{figure}
	\resizebox{\hsize}{!}{\includegraphics{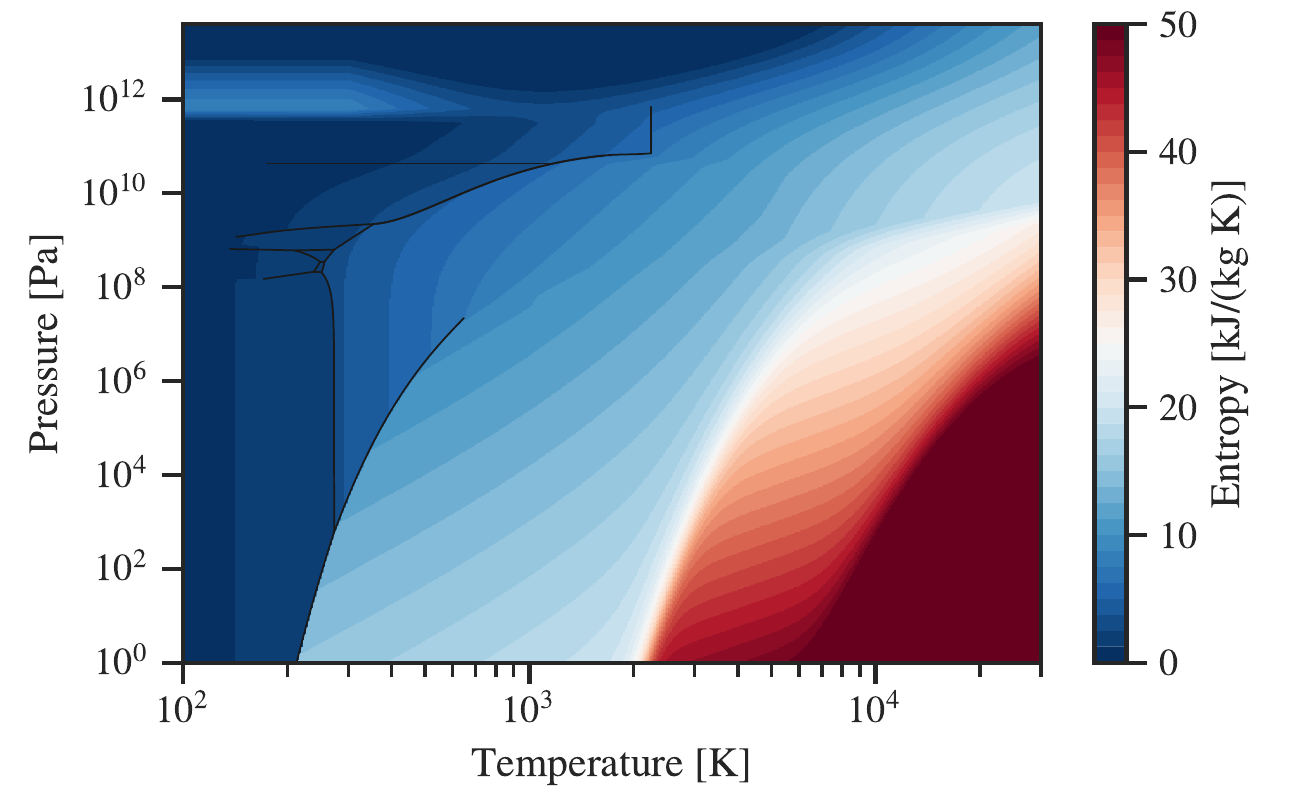}}
	\resizebox{\hsize}{!}{\includegraphics{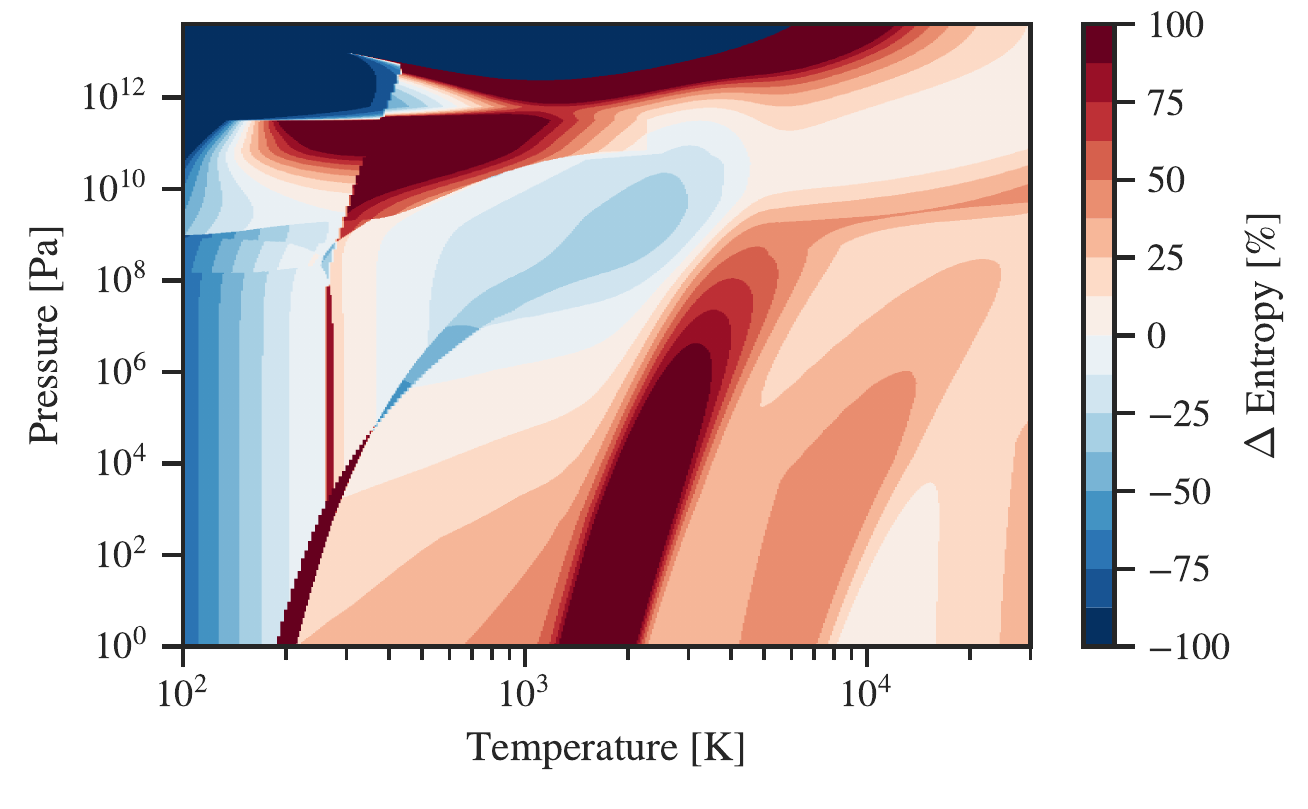}}
	\resizebox{\hsize}{!}{\includegraphics{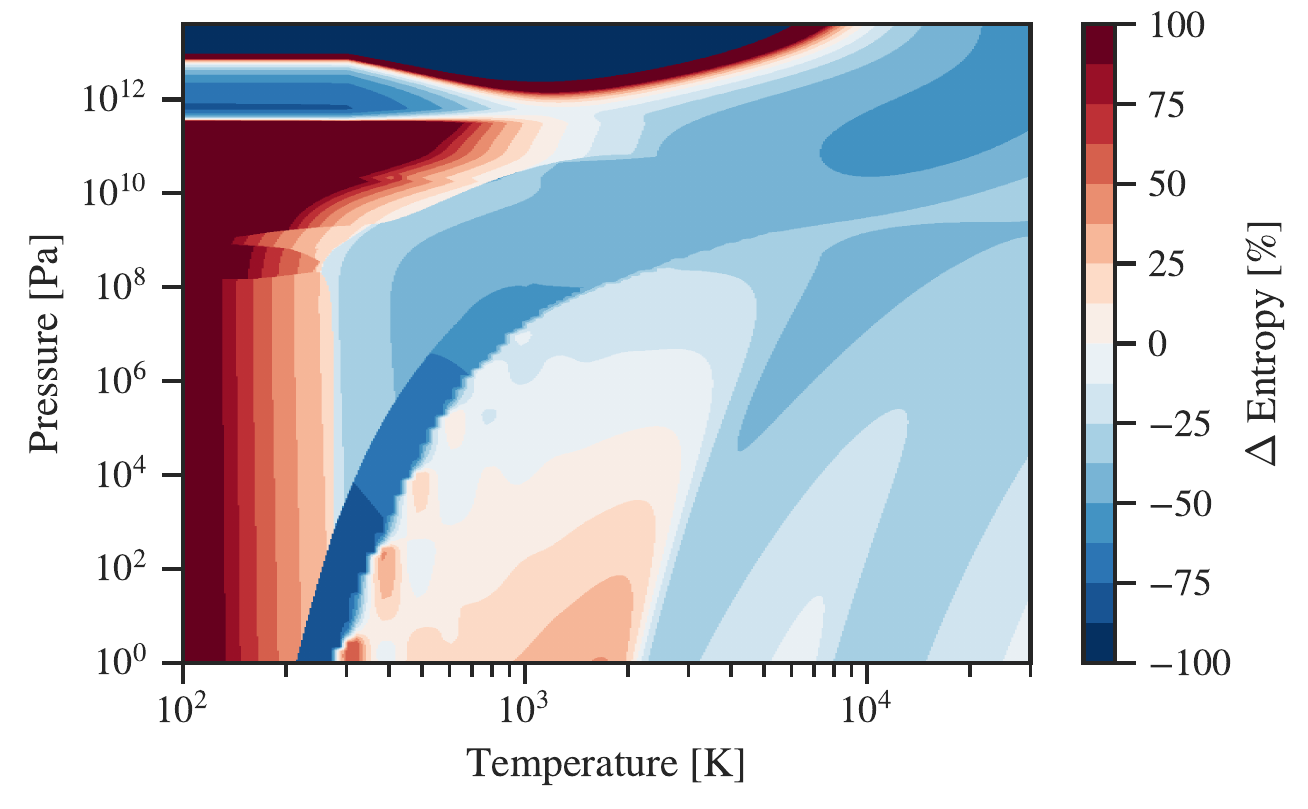}}
	\caption{In the top panel the specific entropy of AQUA as a function of pressure and temperature is shown. In the middle panel we show the relative difference between the specific entropy of ANEOS vs. AQUA. While in the bottom panel the same comparison is performed between QEOS and AQUA.} 
	\label{Fig:WaterEOS_entropy}
\end{figure}

\begin{figure}
	\resizebox{\hsize}{!}{\includegraphics{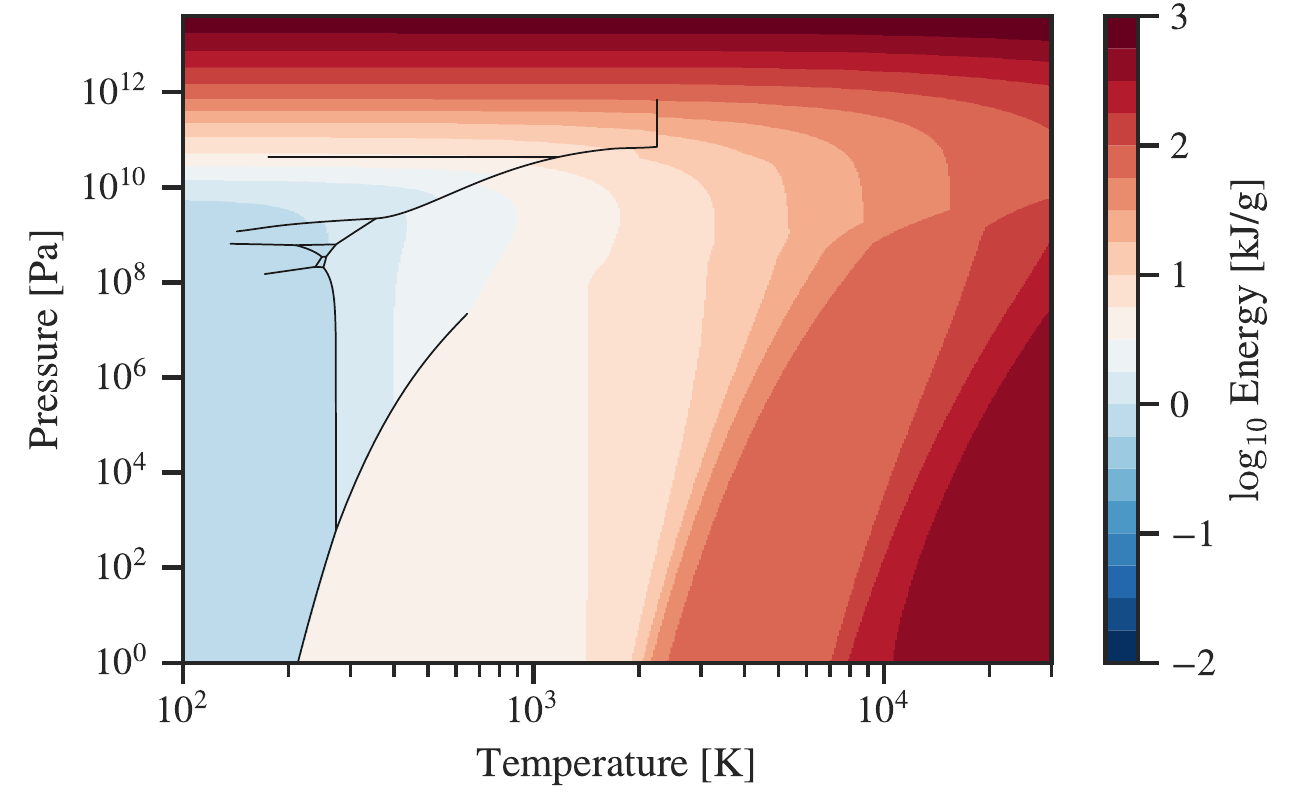}}
	\resizebox{\hsize}{!}{\includegraphics{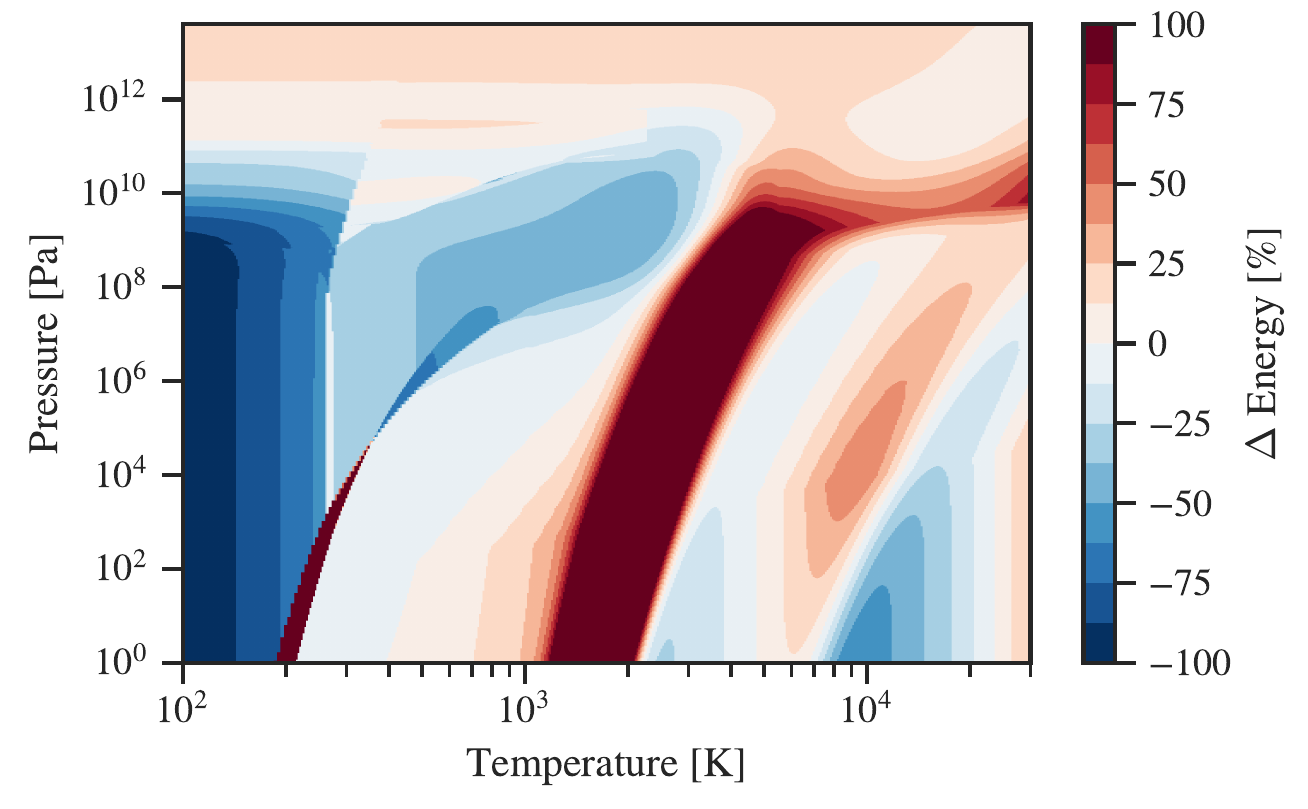}}
	\resizebox{\hsize}{!}{\includegraphics{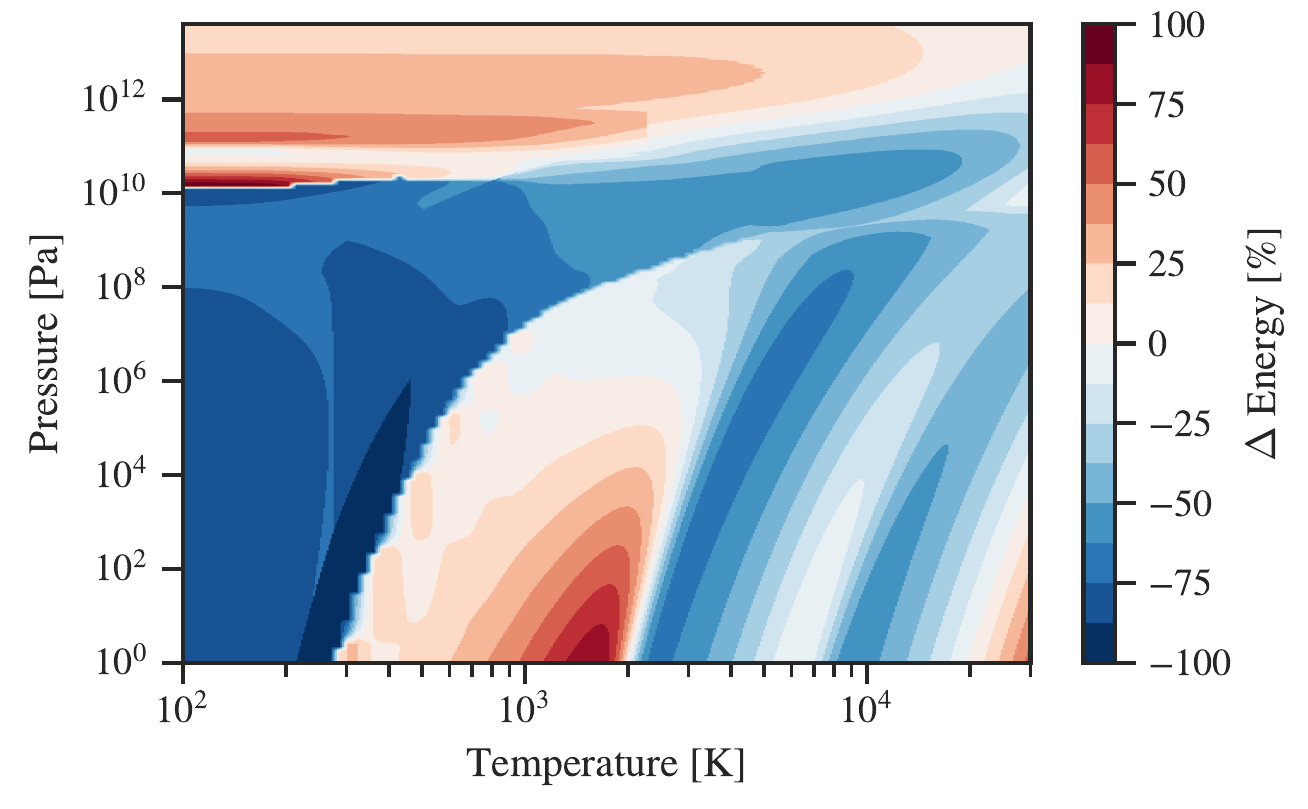}}
	\caption{In the top panel the specific internal energy of AQUA as a function of pressure and temperature is shown. In the middle panel we show the relative difference between the internal energy of ANEOS vs. AQUA. While in the bottom panel the same comparison is performed between QEOS and AQUA. The internal energy potential of QEOS seems to be globally shifted compared to ANEOS and AQUA. Probably due to a different choice of reference state. For the comparison we therefore subtracted 37.5 kJ/g from u(T,P)$_\text{QEOS}$.}
	\label{Fig:WaterEOS_energy}
\end{figure}

As with the other variables we compare the results of the entropy and internal energy calculations with predictions by ANEOS and QEOS. In Fig. \ref{Fig:WaterEOS_entropy}  and \ref{Fig:WaterEOS_energy} we show in the top panel the entropy and internal energy predictions by AQUA as a function of $P$ and $T$. While in the middle panel we show the relative differences compared to ANEOS and in the bottom panel compared to QEOS. The differences are calculated in the same way as for the density in Eq. (\ref{EQ:diff}). Compared to ANEOS the largest differences occur in the region where H$_2$O dissociates. Both entropy and energy differ in this region by a factor of two. For the other part in P--T space the results for the internal energy do not differ more than $\pm 25$ \%. Except a small region in the high pressure ice phases. Contrary the entropy of ANEOS is significantly higher in the region of the high pressure ices between $10^{10}$ and $10^{12}$ Pa. Likely due to the fact that the location of the melting curve in ANEOS is at much lower temperature than the one of AQUA.
	
While the energies and entropies of ANEOS were always within the same order of magnitude. The predictions of QEOS seem to be globally shifted. Compared to AQUA the energies are in average $\sim$ 37.5 kJ/g larger while the entropies are $\sim$ 2 J/(g K) smaller. This shift likely originates from a different choice of reference state. Though since this information is not provided in \citet{vazan_helled_2013}, we can not be certain. Hence for most part of P--T space, the entropies of QEOS are  50\%-75\% smaller than the ones of AQUA. Only in the low temperature vapour region the entropy is a few percent larger. Regarding the internal energies, a shift of 37.5 kJ/g means that for pressures below $\sim10^{10}$ Pa and temperature below $\sim 2000$ K, the predictions of AQUA and QEOS differ by multiple orders of magnitude. Assuming that this shift is due to a different reference state we show in Fig. \ref{Fig:WaterEOS_energy} the difference in energy if the internal energy potential of QEOS would be 37.5 kJ/g smaller. We see that compared to ANEOS the spread in differences is larger. Some differences can be attributed to the shifted vapour curve. While we do not see a strong effect of the melting curve as we do with ANEOS. Though the energy of ice VI is notably smaller than the energy of ice VII and the other low pressure ices. Due to the applied shift in energy, the differences especially at low pressures can not be accurately determined.

\subsection{Bulk Speed of Sound $w(P,T)$}
\begin{figure}
	\resizebox{\hsize}{!}{\includegraphics{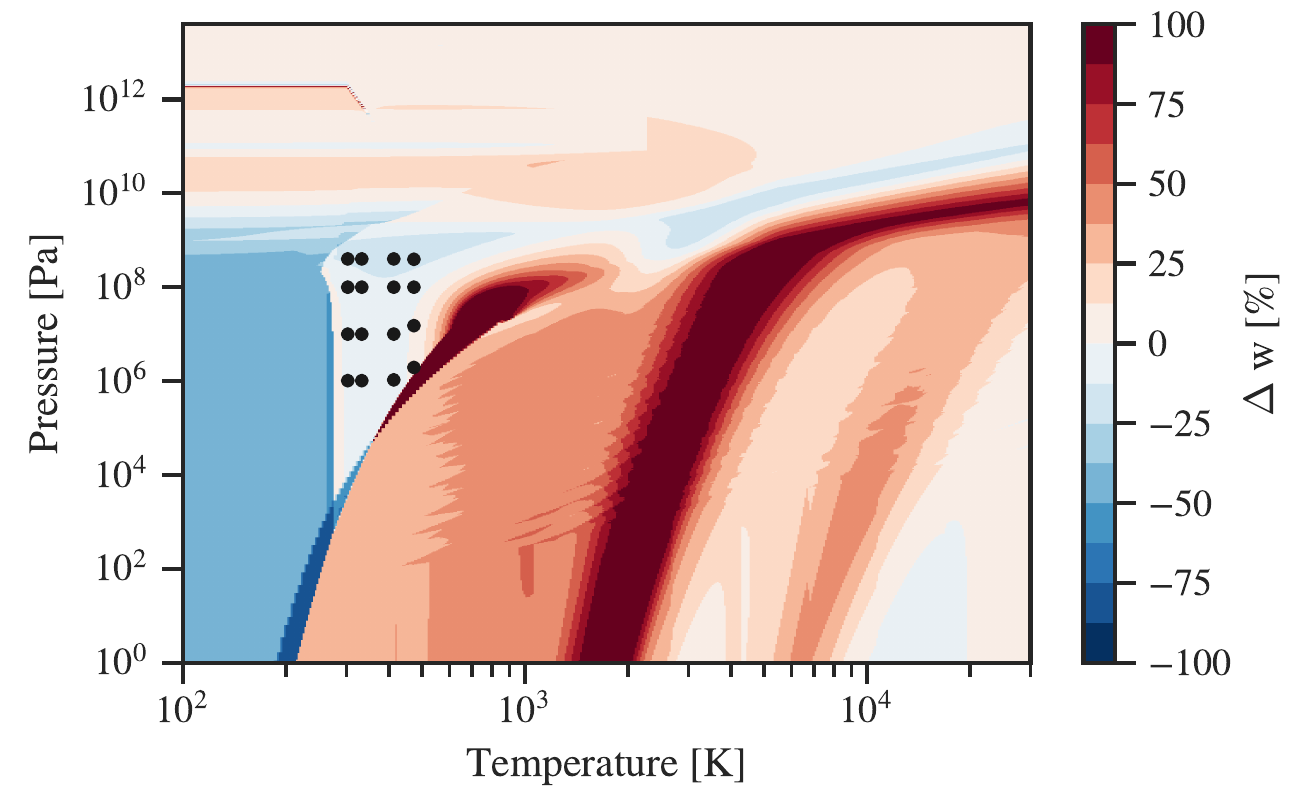}}
	\caption{Relative difference between the bulk speed of sound of ANEOS and AQUA as a function of pressure and temperature. The black dots indicate the location of the experimental data of \citet{lin_speed_2012} shown in Fig. \ref{Fig:Bulk_Speed_Comp}.}
	\label{Fig:Bulk_Speed_Diff}
\end{figure}
\begin{figure}

	\resizebox{\hsize}{!}{\includegraphics{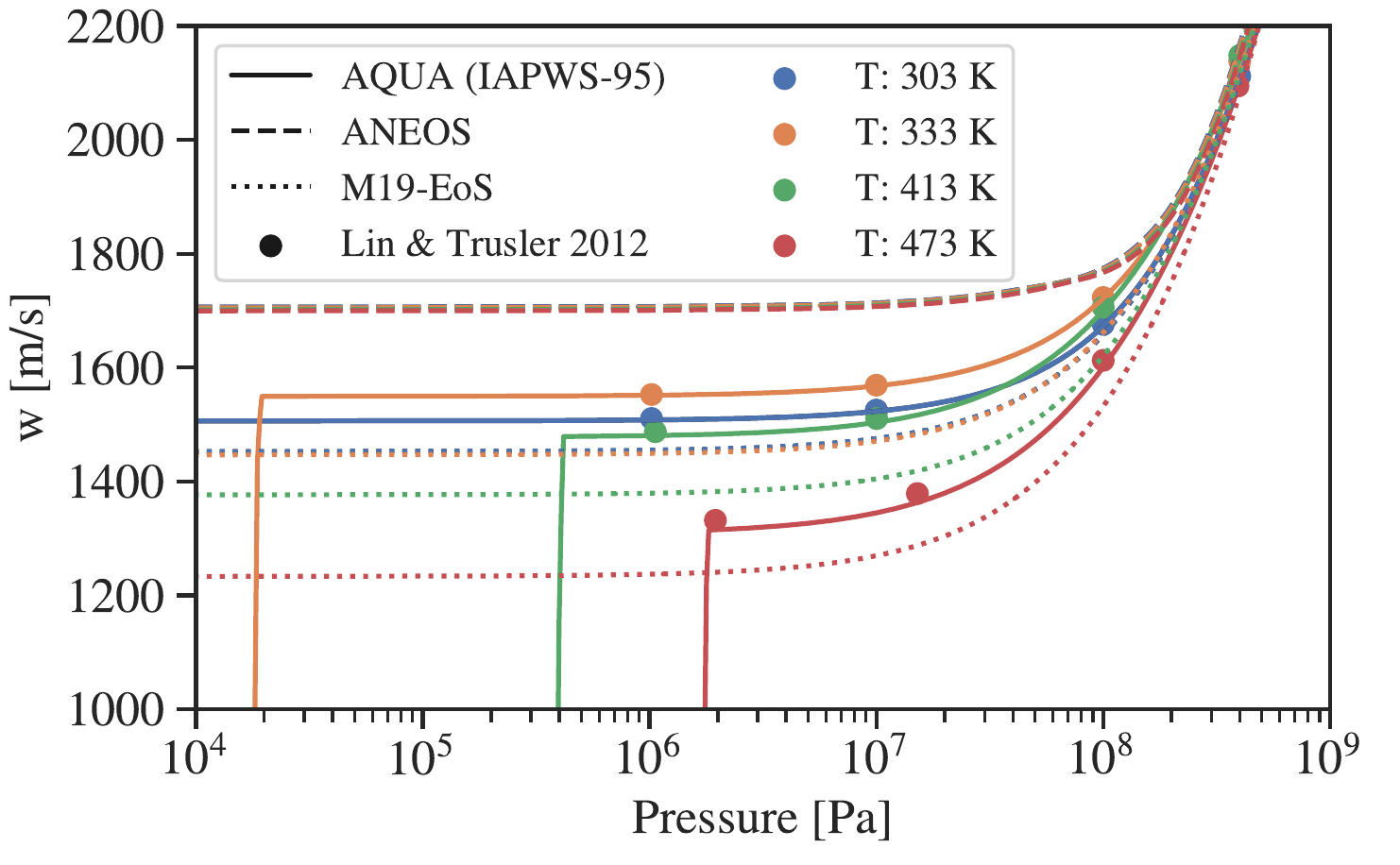}}
	\caption{Comparison of the bulk speed of sound, between AQUA-EoS (solid), ANEOS (dashed), the pure M19-EoS (dotted) and experimental results of \citet{lin_speed_2012}. We would like to point out that the compared range corresponds to region 4 and 5. Hence AQUA-EoS evaluates mainly the IAPWS-95 EoS and above $10^8$ Pa the EoS of \citet{brown_local_2018} is used.}
	\label{Fig:Bulk_Speed_Comp}
\end{figure}
At last we show the results for the bulk speed of sound. Since QEOS of \citet{vazan_helled_2013} does not provide the bulk speed of sound we will compare in Fig. \ref{Fig:Bulk_Speed_Diff} only against ANEOS. Though in Fig. \ref{Fig:Bulk_Speed_Comp} we also show a comparison against experimental results of \citet{lin_speed_2012} at low temperatures. Compared to ANEOS the bulk speed of sound is for most parts within $\pm 40$\%. Notable differences occur throughout the dissociation region, around the critical point and within the region of ice-Ih. At high pressures (>$10^{10}$ Pa) both ANEOS and AQUA results are within 10 \%.

In Fig. \ref{Fig:Bulk_Speed_Comp} one can see that due to the use of the IAPWS-95 EoS the bulk speed of sound of AQUA (solid lines) fits very well the experimental data of \citet{lin_speed_2012}. While ANEOS (dashed lines) over predicts the speed of sound at pressures below $10^{8}$ Pa. Also ANEOS does not show a drop in speed of sound at the vapour curve. For comparison we also show what the pure M19-EoS would predict (dotted lines). Like ANEOS it shows no drop at the vapour curve while predicting a generally lower speed of sound than AQUA.

\section{Effect on the Mass Radius Relation of Planets}
\label{Sec:MR}
We see the main application of the AQUA-EoS in the calculation of internal structures of planets, exoplanets and their moons. To test the effect of different EoS onto these calculations, we determine the mass radius relation for pure water spheres. As already \citet{mazevet_2019} stated, this is a purely academic exercise, but it is still useful since the results solely depend on the used EoS. In Appendix \ref{App:Structuremodel} we explain how we calculate the internal structure of a pure water sphere of given mass and determine its radius. We compare the AQUA-EoS against ANEOS, QEOS, the H$_2$O EoS used in \citet{sotin_massradius_2007} and the mass radius results of \citet{zeng_growth_2019}. \citet{zeng_growth_2019} use a similar selection of EoS to the one proposed in this work i.e. \citet{wagner_iapws_2002}, \citet{frank_constraining_2004}, \citet{french_equation_2009} and \citet{french_redmer_2015} but do not provide a public available EoS. Though we will use their results as a benchmark for our mass radius calculations. A more simple approach was chosen by \citet{sotin_massradius_2007}, which on the other hand use two 3rd order Birch-Murnaghan EoS: one isothermal for the liquid layer and one including temperature corrections for the high pressure ices.
\begin{figure*}
	\begin{minipage}[t]{\hsize}
		\centering
		\includegraphics[width=0.85\linewidth]{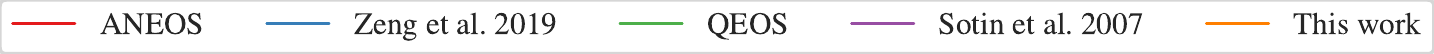}
	\end{minipage}
	\begin{minipage}[t]{0.5\hsize}
		\includegraphics[width=\linewidth]{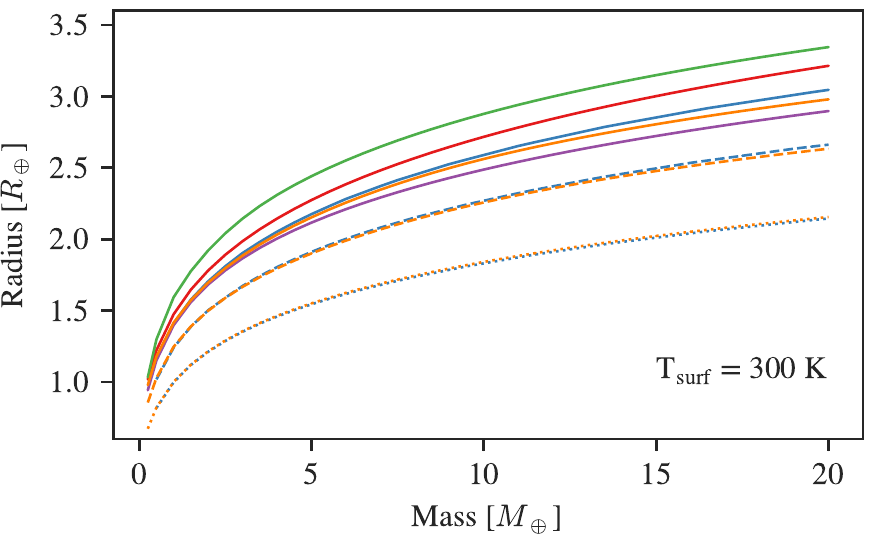}
	\end{minipage}
	\begin{minipage}[t]{0.5\hsize}
		\includegraphics[width=\linewidth]{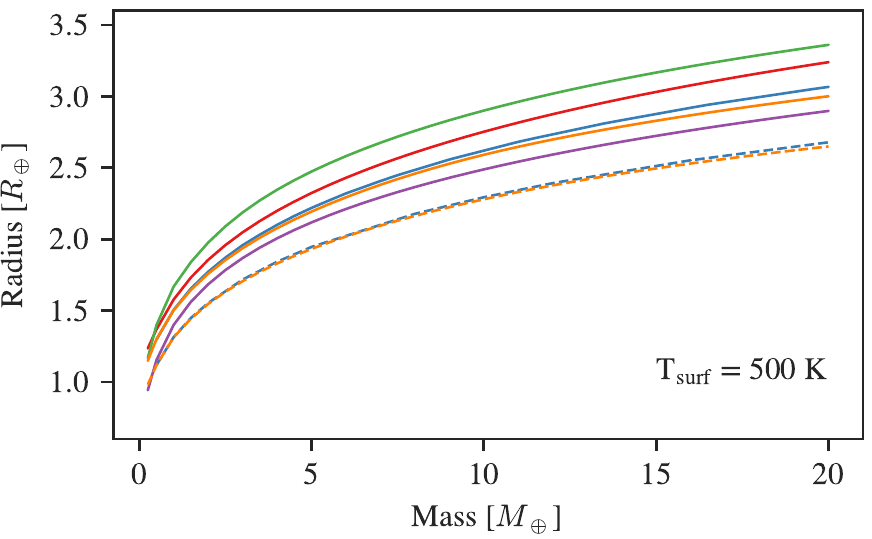}
	\end{minipage}
	\begin{minipage}[t]{0.5\hsize}
		\includegraphics[width=\linewidth]{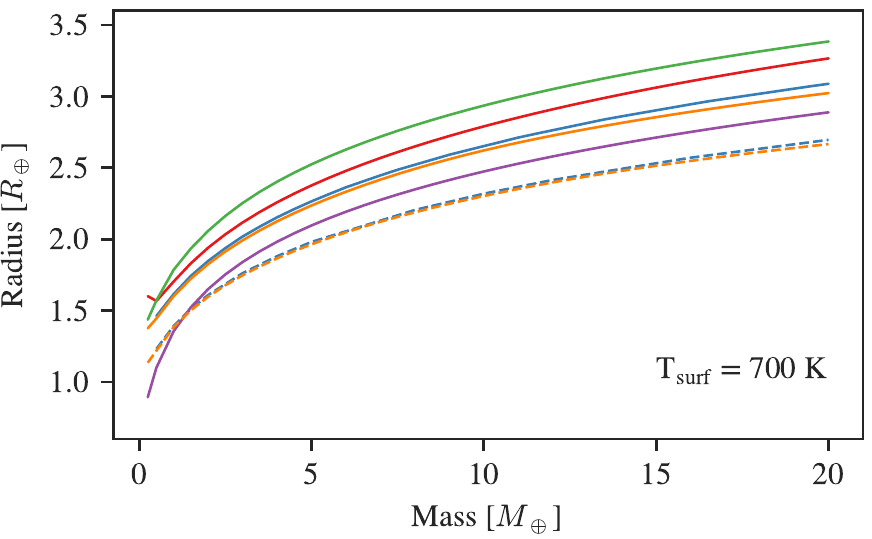}
	\end{minipage}
	\begin{minipage}[t]{0.5\hsize}
		\includegraphics[width=\linewidth]{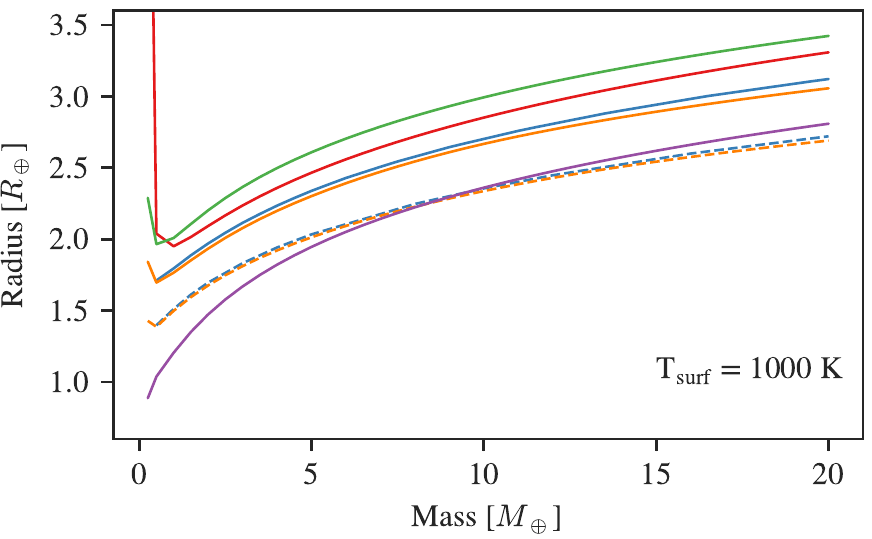}
	\end{minipage}
	\caption{The mass radius relations of isothermal spheres in hydrostatic equilibrium made of 100 wt\% H$_2$O (solid lines) or 50 wt\% H$_2$O and 50 wt\% Earth like composition as in \citet{zeng_growth_2019} (dashed lines), for different EoS and different surface temperatures T$_\text{surf}$. The surface pressure was chosen to be 1 mbar as in \citet{zeng_growth_2019}. For the cases with Earth like composition we used \citet{hakim_new_2018} as EoS for the Fe and \citet{sotin_massradius_2007} to calculate the density in the MgSiO$_3$ layer. In the top left panel also the Earth like composition case \textbf{is} plotted (dotted lines) in order to quantify the contribution from the underlying rocky part to the cases of mixed composition.} 
	\label{Fig:MR}
\end{figure*}
In Fig. \ref{Fig:MR} we show the result of the structure calculations for isothermal water spheres with masses between 0.25 and 20 $M_\oplus$. As in \citet{zeng_growth_2019} we fixed the surface pressure to 1 mbar, while each panel shows the results for a different surface temperature between 300 K and 1000 K.
 
We see that the choice of EoS has a strong effect on the radius for a given water mass. For ANEOS and QEOS one can predict that for large water mass fractions the radii for a given mass will be bigger, due to the lower density at high pressures. This feature is visible in all panels. The results of ANEOS are closer to AQUA, than the ones of QEOS. For both the change in surface temperature does not strongly affect the relative differences compared to AQUA. Contrary the EoS used in \citet{sotin_massradius_2007} shows a bigger difference towards higher temperatures. This is due to the isothermal liquid layer and the absence of a vapour description in \cite{sotin_massradius_2007}. But at 300 K the results only differ by -0.8\% to -3.69\% for \cite{sotin_massradius_2007}.
	
We report that the mass radius relation of \citet{zeng_growth_2019} does predict very similar radii, within $\pm 2.5\%$. Except for a few low mass cases, larger radii are predicted. This is likely due to the fact, that they do not compute a fully isothermal profile but follow the melting curve of the high pressure ice/super-ionic phase, as soon as the isotherm would intersect the melting curve. Which would lead to lower densities at high pressures.

The particular kink in the various mass radius relation at high surface temperatures and low water masses originates from the fact that there is not enough mass to create a steep enough pressure profile and for high enough temperatures the water sphere is then almost completely in the vapour phase, which results in inflated radii. This effect would be much more pronounced if an adiabatic temperature gradient was used, where even for lower surface temperatures the temperature profile would not cross the vapour curve.

For comparison we also plotted in dashed lines the mass radius relation for spheres with a 50 wt\% H$_2$O and 50 wt\% Earth like composition (i.e. 33.75 wt\% MgSiO$_3$ and 16.25 wt\% Fe), as in \citet{zeng_growth_2019}. The results are also listed in Table \ref{Tab:mr_table_rock} in the appendix. For the EoS of the Fe core we used \citet{hakim_new_2018} and \citet{sotin_massradius_2007} for the MgSiO$_3$ layer. In the 300 K panel we show as dotted lines also the pure Earth like composition case, in order to show that any difference in the 50 wt\% H$_2$O case stems from the H$_2$O EoS. The difference in radius for the 50 wt\% H$_2$O case are about a factor two smaller than the difference in the pure H$_2$O case, i.e. between -1\% and 1.1\% of relative difference.

\subsection{Adiabatic Temperature Gradient vs. Isothermal}
For planets with significant amounts of volatile elements the proper treatment of thermal transport is of big importance for the mass radius calculation. We show here the effect of having a fully adiabatic temperature profile instead of an isothermal one, as it was assumed in the last section. 

Adiabatic processes follow an isentrope, as indicated by the dashed lines in Fig. \ref{Fig:WaterEOS_rho}. Hence if one starts below the vapour curve in the gas phase, the adiabat will never cross the vapour curve or one of the melting curves before the water becomes a supercritical fluid. But if we follow an isotherm from the same starting point, we reach the liquid phase at comparable low pressures where in the adiabatic case we would still be in the vapour phase. The adiabatic case resembles a post greenhouse state where all the water is fully mixed in the atmosphere.

\begin{figure}
	\resizebox{\hsize}{!}{\includegraphics{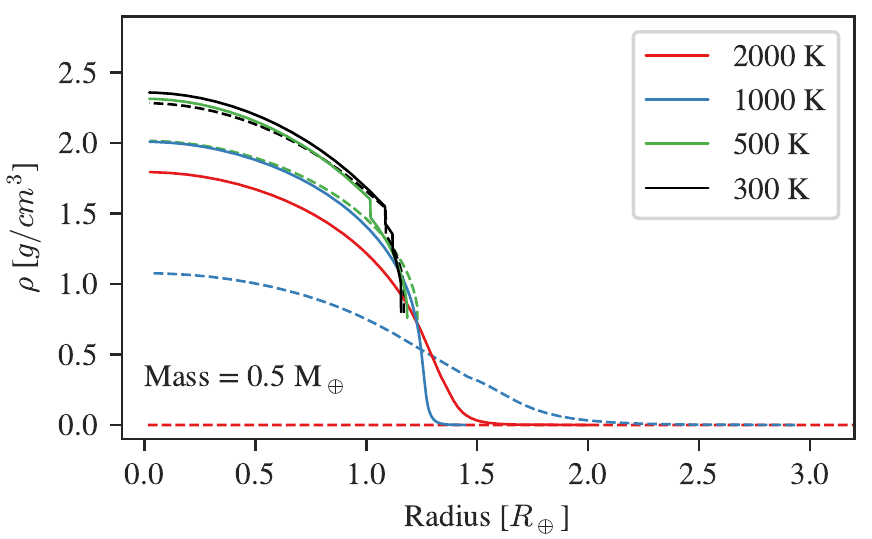}}
	\resizebox{\hsize}{!}{\includegraphics{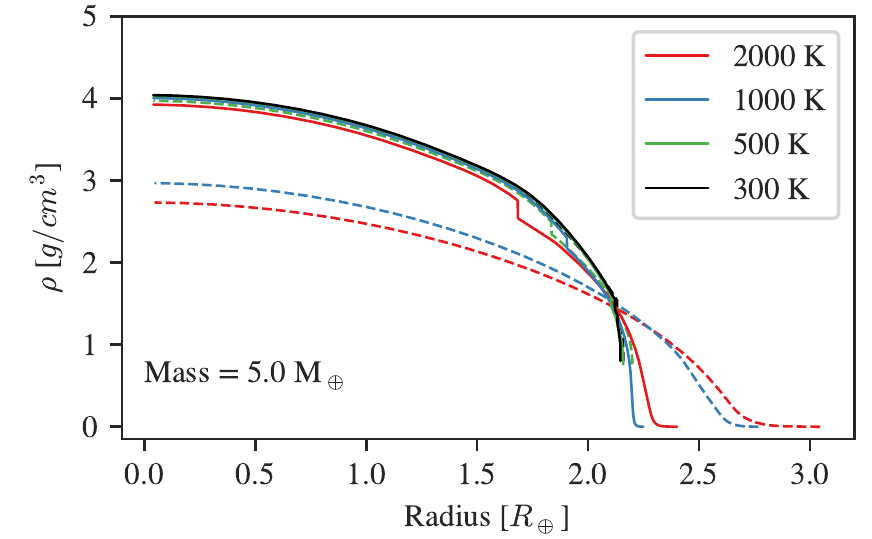}}
	\caption{Density profiles of a water sphere using either an isothermal temperature profile (solid) or an adiabatic temperature gradient (dashed). The surface temperatures are set to four different values (black = 300 K, green = 500 K, blue = 1000 K, red = 2000 K). Following a similar comparison from \citet{mazevet_2019}, we set the surface pressure to the corresponding pressure on the vapour curve. Unless the surface temperature is above the critical point, where the surface pressure is set to 1 bar. In the bottom panel the black dashed curve is overlapped by the solid black and green curves.}
	\label{Fig:Radius_rho}
\end{figure}

In order to quantify this difference we choose as in Sect. 5.1 of \citet{mazevet_2019} two masses of $0.5$ M$_\oplus$ and $5$ M$_\oplus$. For various surface temperatures we calculate the structure of a pure H$_2$O sphere. We use the same model as in the last section, although we set the surface pressure to either the value of the vapour curve or 1 bar if above the critical temperature (as in \citet{mazevet_2019}). We show in Fig. \ref{Fig:Radius_rho} the density as a function of radius for said two masses. We see that considering the two thermal structures, the density profile is considerably different at large surface temperatures (T$_\text{surf}$ > 1000 K), while at low surface temperatures (T$_\text{surf}$ < 300 K) it is almost equal. In the adiabatic case starting at 2000 K it is even below $2\cdot 10^{-4}$ g/cm$^3$ throughout the structure. Here we see the effect of increased adiabatic temperature gradient in the vapour phase compared to the liquid or solid phases. This effect is reduced if the water mass becomes larger and hence the sampled pressure scale increases simultaneously. In Table (\ref{Tab:iso_vs_adiabat}) we list the total radii of all cases shown in Fig. \ref{Fig:Radius_rho}.

One has to remember that these results are based on calculations of pure water spheres, any addition of dense material will cause a steeper pressure profile and hence a more compact radius. We still conclude that the choice of thermal transport has a significant effect on the mass radius relation of a volatile layer made out of H$_2$O.
\begin{table}
	\centering
	\caption{Tabulated radii of a pure water spheres using either an adiabatic temperature gradient (Radius$_\text{Ad}$) or an isothermal temperature profile (Radius$_\text{i.t.s}$) and various surface temperatures. The surface pressure was chosen as in section 5.1 of \citet{mazevet_2019}, i.e., either along the vapour curve or if the temperature was supercritical fixed at 1 bar.}
	\label{Tab:iso_vs_adiabat}
	\begin{tabular}{c c c c}
		\hline
		\hline
		$T_\text{surf}$ [K] & Radius$_\text{Ad}$ [R$_\oplus$] & Radius$_\text{i.t.}$ [R$_\oplus$] & $\delta_R$ [\%] \\
		\hline
		&\multicolumn{2}{c}{Mass = 0.5 M$_\oplus$}&\\
		\hline
		200.0 & 1.154 & 1.151 & 0.27 \\
		300.0 & 1.170 & 1.157 & 1.10 \\
		500.0 & 1.231 & 1.185 & 3.87 \\
		1000.0 & 2.933 & 1.446 & 102.8 \\
		2000.0 & 27.176 & 2.007 & 1254.1 \\
		\hline
		&\multicolumn{2}{c}{Mass = 5.0 M$_\oplus$}&\\
		\hline
		200.0 & 2.148 & 2.143 & 0.22 \\
		300.0 & 2.159 & 2.146 & 0.61 \\
		500.0 & 2.200 & 2.159 & 1.91 \\
		1000.0 & 2.768 & 2.249 & 23.1 \\
		2000.0 & 3.046 & 2.401 & 26.9 \\	
		\hline
	\end{tabular}
\end{table}

\section{Conclusions}
\label{Sec:Conclusions}
We combined the H$_2$O-EoS from \citet{mazevet_2019} with the EoS of \citet{feistel_new_2006,journaux_holistic_2020} and \citet{french_redmer_2015} to include the description of ice phases at low, intermediate and high pressures. For a proper treatment of the liquid phase and gas phase at low pressures we added the EoS by \citet{brown_local_2018,wagner_iapws_2002} and the CEA package \citep{cea1_1994,cea2_1996} for the high temperature low pressure region. This resulted in the tabulated AQUA-EoS (which is available at \url{https://github.com/mnijh/aqua}) providing data for the density $\rho$, adiabatic temperature gradient $\nabla_\text{Ad}$, specific entropy $s$, specific internal energy $u$ and bulk speed of sound $w$. As well as mean molecular weight $\mu$, ionisation fraction $x_i$ and dissociation fraction $x_d$ for a limited region.
The AQUA-EoS offers a multi phase description of all major phases of H$_2$O useful to model the interiors of planets and exoplanets. We recommend the AQUA-EoS for use cases where thermodynamic data over a large range of pressures and temperatures is needed. Though, by its construction, AQUA-EoS is not fully thermodynamical consistent since it is not calculated from a single energy potential, but consistency is sufficient for a large part in P--T space. Nevertheless we remind the reader again, that we do not intend to offer a more accurate description than any EoS tailored to a limited region in P--T space but rather an EoS valid over a larger range of thermodynamical values.

We compared the values of the thermodynamics variables derived from the AQUA-EoS against the values from ANEOS \citep{thompson_aneos_1990} and QEOS \citep{vazan_helled_2013,vazan_explaining_2020}.  
Compared to ANEOS and QEOS, AQUA shows a larger density at P >10 GPa, an effect which is already present in the original M19-EoS. At lower pressures the largest difference are seen in the region of ice-Ih and in the gas below 2000 K.
For $\nabla_\text{Ad}$ the results are more similar though not identical. For the entropy $s$ and also the internal energy $u$, ANEOS predicts higher values by a factor two throughout the dissociation region and along the melting curve of ice-Ih. While in most other regions $s$ and $u$ only differs by $\sim25\%$ compared to AQUA. Except within the ice-VII and ice-X region where the entropy is larger by a factor two given a vastly colder melting curve of ice-VII/X in ANEOS. QEOS shows in average a $\sim 2$ J/(g K) lower entropy than AQUA. Also the internal energy potential $u$ of QEOS seems shifted by 37.5 kJ/g. Given that it is unclear if this shifts stems from a different reference point, the comparison of $s$ and $u$ to AQUA are likely not very accurate. For the bulk speed of sound $w$ we compared against ANEOS and experimental values of \citet{lin_speed_2012}, since QEOS does not provide this thermodynamic quantity.  ANEOS shows the largest differences at pressures below $10^9$ Pa. While given the use of the IAPWS-95 release, AQUA agrees very well with the results of \citet{lin_speed_2012}. 

We further studied the effect of different EoS on the mass radius relation of pure H$_2$O spheres. Within  $\pm 2.5\%$ we reproduce the values of \citet{zeng_growth_2019} which use a similar selection of EoS. The other tested EoS (ANEOS, QEOS and \citet{sotin_massradius_2007}) show much bigger deviations from the radii we calculated. Deviations are between 3\% and 8 \% for ANEOS and between 7\% and 14\% for QEOS, excluding the low water mass cases ($\le 0.5$ M$_\oplus$) where the differences for high temperatures can be larger than 10\%. The H$_2$O EoS of \citet{sotin_massradius_2007} is mainly suited for low surface temperatures since it does not incorporate any vapour phase. For surface temperatures around 300 K it consistently predicts a smaller radius for a given mass by -0.8\% to -3.6\%. Even though we focused in this part on isothermal structures of pure water spheres, which is a mere theoretical test. The differences between EoS are still significant, especially in the view of improved radius estimates from upcoming space based telescopes such as CHEOPS \citep{benz_cheops_2017} or PLATO \citep{rauer_space_2018}. Future work will be needed to unify the description of the thermodynamic properties of water over a wide range of pressure and temperatures.

In a last part we showed that the effect of surface temperature on the total radius is much bigger when we assume an adiabatic temperature profile instead of an isothermal one. This emphasises the importance of a proper treatment of the thermal part of the used EoS when modelling the structure of volatile rich planets.

\begin{acknowledgements}
We thank the anonymous referee for their valuable comments.
We also thank Julia Venturini for the help implementing the CEA code and for various fruitful discussions.  Further, we would like to thank Allona Vazan to provide to us her updated QEOS table for H$_2$O.
J.H. acknowledges the support from the Swiss National Science Foundation under grant 200020\_172746 and 200020\_19203. C.M. acknowledges the support from the Swiss National Science Foundation from grant BSSGI0$\_$155816 ``PlanetsInTime''.

\\
\\

\textit{Software.} For this publication the following software packages have been used: \href{https://docs.python.org/3}{Python 3.6,} CEA (Chemical Equilibrium with Applications) \citep{cea1_1994,cea2_1996}, \href{https://iapws.readthedocs.io/}{Python-iapws,} \href{https://numpy.org/}{Python-numpy,}  \href{https://matplotlib.org/}{Python-matplotlib,}
\href{https://scipy.org/}{Python-scipy,}  \href{https://seaborn.pydata.org/}{Python-seaborn,} \href{https://github.com/Bjournaux/SeaFreeze}{Python-seafreeze.}\citep{journaux_holistic_2020}
\end{acknowledgements}
\bibliographystyle{aa} 
\bibliography{waterpaper.bib} 
\appendix
\section{Derivation of the thermodynamic consistency measure}\label{Sec:app_th_cons}
Unlike other authors which use $\rho$ and $T$ as natural variables for their EoS we choose to use $P$ and $T$ instead. Therefore the thermodynamic consistency measure used e.g. in \citet{becker_ab_2014} needs to be reformulated for the use of $P$ and $T$ as natural variables.
We start with the fundamental thermodynamic relation in terms of the internal energy U, i.e.
\begin{equation}
dU(T,P) = TdS(T,P) - PdV(T,P).
\end{equation}
Then we replace the differentials of the internal energy, entropy and the volume with the following relations
\begin{align}
	dS(T,P) &= \left(\frac{\partial S(T,P)}{\partial T}\right)_P dT + \left(\frac{\partial S(T,P)}{\partial P}\right)_T dP\\
	dV(T,P) &= \left(\frac{\partial V(T,P)}{\partial T}\right)_P dT + \left(\frac{\partial V(T,P)}{\partial P}\right)_T dP	\\
	dU(T,P) &= \left(\frac{\partial U(T,P)}{\partial T}\right)_P dT + \left(\frac{\partial U(T,P)}{\partial P}\right)_T dP	
\end{align}
and sort the pressure and temperature derivatives to one side of the equation each
\begin{equation}
\begin{split}
	\left(\frac{\partial U(T,P)}{\partial T}\right)_P dT -T\left(\frac{\partial S(T,P)}{\partial T}\right)_P dT + P\left(\frac{\partial V(T,P)}{\partial T}\right)_P dT  =\\ -\left(\frac{\partial U(T,P)}{\partial P}\right)_T dP + T\left(\frac{\partial S(T,P)}{\partial P}\right)_T dP - P\left(\frac{\partial V(T,P)}{\partial P}\right)_T dP.\label{Eq:app_du1}
\end{split}	
\end{equation}
Using Bridgman's thermodynamic equations \citep{bridgman_complete_1914} we can replace some of the partial derivatives using the following relations
\begin{align}
	\left(\frac{\partial U(T,P)}{\partial T}\right)_P &=C_P-P\left(\frac{\partial V(T,P)}{\partial T}\right)_P \\
	\left(\frac{\partial S(T,P)}{\partial T}\right)_P &=\frac{C_P(T,P)}{T} \\
	\left(\frac{\partial U(T,P)}{\partial P}\right)_T &=-T\left(\frac{\partial V(T,P)}{\partial T}\right)_P-P\left(\frac{\partial V(T,P)}{\partial P}\right)_T.
\end{align}
Hence Eq. (\ref{Eq:app_du1}) can be written as
\begin{equation}
\begin{split}
	0 = T\left(\frac{\partial V(T,P)}{\partial T}\right)_P dP + T\left(\frac{\partial S(T,P)}{\partial P}\right)_T dP .\label{Eq:app_du2}
\end{split}	
\end{equation}
Next, we can divide both sides by $T\cdot dP$ which results in one of the Maxwell relations
 \begin{equation}
 \left(\frac{\partial S(T,P)}{\partial P}\right)_T = -\left(\frac{\partial V(T,P)}{\partial T}\right)_P = \frac{1}{\rho(T,P)^2}\left(\frac{\partial \rho(T,P)}{\partial T}\right)_P . \label{EQ:maxwell}
 \end{equation}
Similar to \citet{becker_ab_2014} we define a measure of thermodynamic consistency, which compares the caloric left hand side of Eq.(\ref{EQ:maxwell}) with the mechanical right hand side:
\begin{equation}
\Delta_{Th.c.} \equiv 1-\frac{\rho(T,P)^2\left(\frac{\partial S(T,P)}{\partial P}\right)_T}{\left(\frac{\partial \rho(T,P)}{\partial T}\right)_P}. \label{Eq:th_cons}
\end{equation}

\section{Structure model}
\label{App:Structuremodel}
To determine the mass radius relation for pure H$_2$O spheres (or 50 wt\% H$_2$O when compared to \citet{zeng_growth_2019}), we solve the mechanical an thermal structure equations in the Lagrangian notation, as in \citet{kippenhahn_stellar_2012} for stellar structures. We assume a constant luminosity throughout the structure, i.e. we neglect potential heat sources within the planet. The remaining structure equations for a static, 1D-spherically symmetric sphere in hydrostatic equilibrium are then given by
\begin{align}
\frac{\partial r}{\partial m} &= \frac{1}{4\pi r^2 \rho},\label{EQ:mass_cons} \\
\frac{\partial P}{\partial m} &= -\frac{Gm}{4\pi r^4}, \label{EQ:hydrostatic_eq}\\
\frac{\partial T}{\partial m} &= \frac{\partial P}{\partial m }\frac{T}{P}\nabla_\text{Ad} \label{EQ:thermal_transport}
\end{align}
where $\nabla_\text{Ad}$ is the adiabatic temperature gradient as defined in Eq. (\ref{EQ:ad_grad}), $r$ is the radius, $m$ is the mass within radius r, P is the pressure and T the temperature. For a given total mass, surface pressure and surface temperature we use a bidirectional shooting method to solve the two point boundary value problem, posed by Eqs. (\ref{EQ:mass_cons})-(\ref{EQ:thermal_transport}). The equations are integrated using a 5th order Cash-Karp Runge-Kutta method, similar to the one described in \citet{nr_press}. From this calculation we get the mechanical and thermal structure as a function of $m$. From which we can extract the total radius at $m=M_\text{tot}$. If not stated differently the surface pressure is set to 1 mbar as in \citet{zeng_growth_2019}, which is a first order approximation of the depth of the transit radius.

At each numerical step in the Runge-Kutta method, the equation of state is evaluated to determine $\rho(P,T)$ and $\nabla_\text{Ad}(P,T)$. For an isothermal structure $\nabla_\text{Ad}(P,T)$ is simply set to zero. As described in the main text we test various water equation of state. In the case where we compare with the results of \citet{zeng_growth_2019} we split the structure into three layers, an iron core using the EoS of \citet{hakim_new_2018} (16.25 wt\%), a silicate mantle as in \citet{sotin_massradius_2007} (33.75 wt\%) and a water layer (50 wt\%). Similar to this work, \citet{zeng_growth_2019} use multiple EoS for the water layer, i.e. \citet{wagner_iapws_2002}, \citet{frank_constraining_2004}, \citet{french_equation_2009} and \citet{french_redmer_2015}.

\begin{table*}[h]
	\caption{Mass radius relation for isothermal pure H$_2$O spheres and various EoS. The surface boundary conditions are $T_\text{Surf} = 300$ K and $P_\text{Surf}=1$ mbar following \citet{zeng_growth_2019}.}
	\label{Tab:mr_table_300K}
	\centering
	\begin{tabular}{c c c c c c c c c c}
		\hline\hline
		 & AQUA &\multicolumn{2}{c}{ANEOS}&\multicolumn{2}{c}{Zeng et al. (2019)} &\multicolumn{2}{c}{QEOS}&\multicolumn{2}{c}{Sotin et al. (2007)} \\ 
		\hline
		Mass [M$_\oplus$] & Radius [R$_\oplus$] & Radius [R$_\oplus$] & $\delta_\text{R}$ [\%] & Radius [R$_\oplus$] & $\delta_\text{R}$ [\%] &Radius [R$_\oplus$] & $\delta_\text{R}$ [\%] &Radius [R$_\oplus$] & $\delta_\text{R}$ [\%] \\
		\hline
		0.10 & 0.768 & 0.819 & 6.58 & 0.726 & -5.52 & 0.763 & -0.62 & 0.719 & -6.36 \\
		0.25 & 0.978 & 1.019 & 4.20 & 0.953 & -2.53 & 1.033 & 5.63 & 0.942 & -3.60 \\
		0.50 & 1.178 & 1.222 & 3.71 & 1.161 & -1.46 & 1.298 & 10.12 & 1.152 & -2.26 \\
		1.00 & 1.416 & 1.474 & 4.15 & 1.41 & -0.42 & 1.594 & 12.58 & 1.398 & -1.25 \\
		1.50 & 1.573 & 1.647 & 4.66 & 1.577 & 0.24 & 1.777 & 12.92 & 1.56 & -0.86 \\
		2.00 & 1.696 & 1.781 & 4.99 & 1.705 & 0.51 & 1.921 & 13.28 & 1.682 & -0.79 \\
		2.50 & 1.798 & 1.891 & 5.20 & 1.811 & 0.71 & 2.042 & 13.56 & 1.782 & -0.88 \\
		3.00 & 1.886 & 1.986 & 5.33 & 1.903 & 0.94 & 2.145 & 13.73 & 1.866 & -1.03 \\
		3.50 & 1.963 & 2.07 & 5.44 & 1.981 & 0.91 & 2.233 & 13.75 & 1.94 & -1.19 \\
		4.00 & 2.032 & 2.145 & 5.52 & 2.053 & 0.99 & 2.31 & 13.67 & 2.005 & -1.37 \\
		4.50 & 2.095 & 2.212 & 5.58 & 2.117 & 1.02 & 2.379 & 13.54 & 2.063 & -1.54 \\
		5.00 & 2.153 & 2.275 & 5.63 & 2.176 & 1.04 & 2.442 & 13.39 & 2.116 & -1.72 \\
		6.00 & 2.257 & 2.385 & 5.70 & 2.282 & 1.13 & 2.552 & 13.09 & 2.21 & -2.05 \\
		7.00 & 2.347 & 2.482 & 5.76 & 2.372 & 1.06 & 2.648 & 12.82 & 2.292 & -2.34 \\
		8.00 & 2.427 & 2.568 & 5.83 & 2.451 & 1.01 & 2.733 & 12.61 & 2.364 & -2.58 \\
		9.00 & 2.498 & 2.646 & 5.92 & 2.526 & 1.11 & 2.809 & 12.44 & 2.429 & -2.77 \\
		10.0 & 2.562 & 2.717 & 6.05 & 2.59 & 1.07 & 2.878 & 12.32 & 2.488 & -2.90 \\
		12.0 & 2.672 & 2.843 & 6.42 & 2.707 & 1.32 & 2.999 & 12.25 & 2.592 & -2.98 \\
		14.0 & 2.764 & 2.952 & 6.82 & 2.809 & 1.62 & 3.102 & 12.25 & 2.682 & -2.96 \\
		16.0 & 2.844 & 3.049 & 7.20 & 2.896 & 1.83 & 3.193 & 12.26 & 2.762 & -2.91 \\
		18.0 & 2.916 & 3.136 & 7.55 & 2.973 & 1.98 & 3.273 & 12.27 & 2.833 & -2.84 \\
		20.0 & 2.98 & 3.214 & 7.88 & 3.046 & 2.22 & 3.346 & 12.28 & 2.898 & -2.75 \\

		\hline
		
		\hline
	\end{tabular}
\end{table*}
\begin{table*}
	\caption{Mass radius relation for isothermal pure H$_2$O spheres and various EoS. The surface boundary conditions are $T_\text{Surf} = 500$ K and $P_\text{Surf}=1$ mbar following \citet{zeng_growth_2019}.}
	\label{Tab:mr_table_500K}
	\centering
	\begin{tabular}{c c c c c c c c c c}
		\hline\hline
		& AQUA &\multicolumn{2}{c}{ANEOS}&\multicolumn{2}{c}{Zeng et al. (2019)} &\multicolumn{2}{c}{QEOS}&\multicolumn{2}{c}{Sotin et al. (2007)} \\ 
		\hline
		Mass [M$_\oplus$] & Radius [R$_\oplus$] & Radius [R$_\oplus$] & $\delta_\text{R}$ [\%] & Radius [R$_\oplus$] & $\delta_\text{R}$ [\%] &Radius [R$_\oplus$] & $\delta_\text{R}$ [\%] &Radius [R$_\oplus$] & $\delta_\text{R}$ [\%] \\
		\hline
		0.10 & 1.062 & 1.253 & 18.04 & 1.065 & 0.28 & 0.971 & -8.52 & 0.717 & -32.45 \\
		0.25 & 1.148 & 1.235 & 7.57 & 1.145 & -0.28 & 1.168 & 1.70 & 0.942 & -17.95 \\
		0.50 & 1.299 & 1.368 & 5.28 & 1.296 & -0.23 & 1.397 & 7.56 & 1.152 & -11.29 \\
		1.00 & 1.502 & 1.577 & 4.99 & 1.505 & 0.16 & 1.668 & 11.03 & 1.399 & -6.85 \\
		1.50 & 1.645 & 1.732 & 5.29 & 1.654 & 0.59 & 1.839 & 11.85 & 1.561 & -5.06 \\
		2.00 & 1.758 & 1.855 & 5.52 & 1.772 & 0.78 & 1.976 & 12.4 & 1.684 & -4.21 \\
		2.50 & 1.854 & 1.958 & 5.65 & 1.871 & 0.92 & 2.09 & 12.75 & 1.784 & -3.78 \\
		3.00 & 1.937 & 2.048 & 5.74 & 1.958 & 1.09 & 2.187 & 12.93 & 1.868 & -3.57 \\
		3.50 & 2.011 & 2.127 & 5.80 & 2.032 & 1.05 & 2.272 & 12.98 & 1.941 & -3.47 \\
		4.00 & 2.077 & 2.199 & 5.85 & 2.10 & 1.11 & 2.346 & 12.95 & 2.006 & -3.43 \\
		4.50 & 2.138 & 2.264 & 5.89 & 2.163 & 1.16 & 2.413 & 12.88 & 2.064 & -3.43 \\
		5.00 & 2.194 & 2.324 & 5.92 & 2.219 & 1.17 & 2.474 & 12.79 & 2.118 & -3.46 \\
		6.00 & 2.294 & 2.43 & 5.96 & 2.321 & 1.19 & 2.582 & 12.57 & 2.212 & -3.57 \\
		7.00 & 2.381 & 2.524 & 6.00 & 2.408 & 1.11 & 2.676 & 12.37 & 2.293 & -3.70 \\
		8.00 & 2.459 & 2.608 & 6.05 & 2.485 & 1.06 & 2.759 & 12.18 & 2.365 & -3.82 \\
		9.00 & 2.529 & 2.684 & 6.12 & 2.558 & 1.16 & 2.833 & 12.04 & 2.43 & -3.90 \\
		10.0 & 2.591 & 2.753 & 6.23 & 2.62 & 1.12 & 2.901 & 11.95 & 2.489 & -3.95 \\
		12.0 & 2.698 & 2.876 & 6.57 & 2.735 & 1.34 & 3.02 & 11.9 & 2.593 & -3.91 \\
		14.0 & 2.789 & 2.983 & 6.95 & 2.833 & 1.60 & 3.121 & 11.92 & 2.683 & -3.80 \\
		16.0 & 2.868 & 3.078 & 7.31 & 2.921 & 1.84 & 3.211 & 11.95 & 2.762 & -3.68 \\
		18.0 & 2.938 & 3.163 & 7.64 & 2.996 & 1.97 & 3.29 & 11.97 & 2.834 & -3.55 \\
		20.0 & 3.001 & 3.24 & 7.95 & 3.067 & 2.18 & 3.361 & 11.99 & 2.898 & -3.43 \\
	\hline
	
	\hline
\end{tabular}
\end{table*}	

\begin{table*}
	\caption{Mass radius relation for isothermal pure H$_2$O spheres and various EoS. The surface boundary conditions are $T_\text{Surf} = 700$ K and $P_\text{Surf}=1$ mbar following \citet{zeng_growth_2019}.}
	\label{Tab:mr_table_700K}
	\centering
	\begin{tabular}{c c c c c c c c c c}
		\hline\hline
		& AQUA &\multicolumn{2}{c}{ANEOS}&\multicolumn{2}{c}{Zeng et al. (2019)} &\multicolumn{2}{c}{QEOS}&\multicolumn{2}{c}{Sotin et al. (2007)} \\ 
		\hline
		Mass [M$_\oplus$] & Radius [R$_\oplus$] & Radius [R$_\oplus$] & $\delta_\text{R}$ [\%] & Radius [R$_\oplus$] & $\delta_\text{R}$ [\%] &Radius [R$_\oplus$] & $\delta_\text{R}$ [\%] &Radius [R$_\oplus$] & $\delta_\text{R}$ [\%] \\
		\hline
		0.25 & 1.376 & 1.601 & 16.36 & 1.382 & 0.45 & 1.437 & 4.43 & 0.894 & -35.05 \\
		0.50 & 1.445 & 1.567 & 8.47 & 1.462 & 1.17 & 1.571 & 8.75 & 1.10 & -23.85 \\
		1.00 & 1.601 & 1.703 & 6.41 & 1.615 & 0.91 & 1.786 & 11.54 & 1.355 & -15.34 \\
		1.50 & 1.724 & 1.831 & 6.26 & 1.743 & 1.13 & 1.935 & 12.27 & 1.523 & -11.62 \\
		2.00 & 1.826 & 1.94 & 6.27 & 1.847 & 1.18 & 2.058 & 12.73 & 1.65 & -9.60 \\
		2.50 & 1.914 & 2.034 & 6.27 & 1.938 & 1.22 & 2.163 & 13.0 & 1.753 & -8.40 \\
		3.00 & 1.992 & 2.117 & 6.27 & 2.019 & 1.34 & 2.253 & 13.11 & 1.84 & -7.63 \\
		3.50 & 2.062 & 2.191 & 6.27 & 2.088 & 1.29 & 2.332 & 13.12 & 1.915 & -7.10 \\
		4.00 & 2.125 & 2.258 & 6.28 & 2.153 & 1.32 & 2.403 & 13.08 & 1.982 & -6.72 \\
		4.50 & 2.183 & 2.32 & 6.28 & 2.211 & 1.30 & 2.466 & 12.99 & 2.042 & -6.45 \\
		5.00 & 2.236 & 2.377 & 6.28 & 2.265 & 1.29 & 2.525 & 12.89 & 2.096 & -6.26 \\
		6.00 & 2.333 & 2.479 & 6.27 & 2.364 & 1.33 & 2.628 & 12.66 & 2.192 & -6.01 \\
		7.00 & 2.418 & 2.569 & 6.27 & 2.447 & 1.23 & 2.718 & 12.44 & 2.275 & -5.87 \\
		8.00 & 2.493 & 2.65 & 6.29 & 2.522 & 1.15 & 2.798 & 12.25 & 2.349 & -5.79 \\
		9.00 & 2.561 & 2.723 & 6.34 & 2.592 & 1.21 & 2.871 & 12.09 & 2.415 & -5.71 \\
		10.0 & 2.622 & 2.79 & 6.43 & 2.652 & 1.16 & 2.936 & 11.98 & 2.474 & -5.62 \\
		12.0 & 2.726 & 2.91 & 6.74 & 2.763 & 1.36 & 3.051 & 11.92 & 2.58 & -5.38 \\
		14.0 & 2.815 & 3.014 & 7.09 & 2.861 & 1.62 & 3.151 & 11.92 & 2.671 & -5.12 \\
		16.0 & 2.892 & 3.107 & 7.42 & 2.945 & 1.82 & 3.238 & 11.93 & 2.751 & -4.88 \\
		18.0 & 2.961 & 3.191 & 7.74 & 3.019 & 1.94 & 3.315 & 11.95 & 2.823 & -4.66 \\
		20.0 & 3.024 & 3.266 & 8.03 & 3.089 & 2.15 & 3.385 & 11.95 & 2.889 & -4.46 \\
		\hline
		
		\hline
	\end{tabular}
\end{table*}

\begin{table*}
	\caption{Mass radius relation for isothermal pure H$_2$O spheres and various EoS. The surface boundary conditions are $T_\text{Surf} = 1000$ K and $P_\text{Surf}=1$ mbar following \citet{zeng_growth_2019}.}
	\label{Tab:mr_table_1000K}
	\centering
	\begin{tabular}{c c c c c c c c c c}
		\hline\hline
		& AQUA &\multicolumn{2}{c}{ANEOS}&\multicolumn{2}{c}{Zeng et al. (2019)} &\multicolumn{2}{c}{QEOS}&\multicolumn{2}{c}{Sotin et al. (2007)} \\ 
		\hline
		Mass [M$_\oplus$] & Radius [R$_\oplus$] & Radius [R$_\oplus$] & $\delta_\text{R}$ [\%] & Radius [R$_\oplus$] & $\delta_\text{R}$ [\%] &Radius [R$_\oplus$] & $\delta_\text{R}$ [\%] &Radius [R$_\oplus$] & $\delta_\text{R}$ [\%] \\
		\hline
		0.25 & 1.842 & 6.42 & 248.45 & 1.866 & 1.28 & 2.29 & 24.27 & 0.887 & -51.86 \\
		0.50 & 1.696 & 2.039 & 20.22 & 1.711 & 0.87 & 1.966 & 15.89 & 1.038 & -38.82 \\
		1.00 & 1.766 & 1.950 & 10.46 & 1.795 & 1.68 & 2.008 & 13.74 & 1.205 & -31.76 \\
		1.50 & 1.852 & 2.015 & 8.82 & 1.887 & 1.90 & 2.106 & 13.74 & 1.350 & -27.09 \\
		2.00 & 1.934 & 2.093 & 8.20 & 1.970 & 1.87 & 2.202 & 13.87 & 1.473 & -23.82 \\
		2.50 & 2.009 & 2.167 & 7.84 & 2.045 & 1.78 & 2.289 & 13.9 & 1.578 & -21.46 \\
		3.00 & 2.078 & 2.236 & 7.60 & 2.115 & 1.76 & 2.366 & 13.85 & 1.669 & -19.68 \\
		3.50 & 2.141 & 2.300 & 7.43 & 2.177 & 1.66 & 2.435 & 13.75 & 1.75 & -18.28 \\
		4.00 & 2.199 & 2.359 & 7.31 & 2.235 & 1.65 & 2.498 & 13.63 & 1.822 & -17.15 \\
		4.50 & 2.252 & 2.414 & 7.21 & 2.288 & 1.61 & 2.556 & 13.49 & 1.887 & -16.22 \\
		5.00 & 2.302 & 2.466 & 7.13 & 2.339 & 1.60 & 2.609 & 13.35 & 1.946 & -15.45 \\
		6.00 & 2.392 & 2.560 & 7.00 & 2.429 & 1.54 & 2.705 & 13.06 & 2.052 & -14.24 \\
		7.00 & 2.473 & 2.643 & 6.91 & 2.507 & 1.40 & 2.789 & 12.78 & 2.143 & -13.33 \\
		8.00 & 2.545 & 2.719 & 6.86 & 2.578 & 1.32 & 2.864 & 12.55 & 2.223 & -12.62 \\
		9.00 & 2.609 & 2.788 & 6.85 & 2.645 & 1.36 & 2.932 & 12.35 & 2.296 & -12.02 \\
		10.0 & 2.668 & 2.852 & 6.90 & 2.702 & 1.29 & 2.994 & 12.21 & 2.361 & -11.5 \\
		12.0 & 2.769 & 2.966 & 7.12 & 2.808 & 1.43 & 3.103 & 12.09 & 2.476 & -10.57 \\
		14.0 & 2.854 & 3.066 & 7.41 & 2.901 & 1.64 & 3.198 & 12.05 & 2.575 & -9.80 \\
		16.0 & 2.929 & 3.155 & 7.70 & 2.983 & 1.82 & 3.282 & 12.03 & 2.661 & -9.15 \\
		18.0 & 2.996 & 3.235 & 7.98 & 3.054 & 1.93 & 3.356 & 12.01 & 2.739 & -8.59 \\
		20.0 & 3.057 & 3.309 & 8.24 & 3.122 & 2.12 & 3.424 & 12.0 & 2.809 & -8.12 \\
		\hline
		
		\hline
	\end{tabular}
\end{table*}

\begin{table*}
	\caption{Mass radius relation for isothermal spheres with 50 wt\% H$_2$O, 33.75 wt\% MgSiO$_3$ and 16.25 wt\% Fe. Radius$_\text{A}$ was calculated using AQUA-EoS for the H$_2$O, the EoS from \citet{sotin_massradius_2007} for the MgSiO3 and the EoS from \citet{hakim_new_2018} for Fe. Radius$_\textbf{Z}$ is based on the results from \citet{zeng_growth_2019}. The surface pressure is $P_\text{Surf}=1$ mBar.}
	\label{Tab:mr_table_rock}
	\centering
	\begin{tabular}{c c c c c c c}
		\hline\hline
		T$_\text{surf}$& \multicolumn{3}{c}{300 K}  &\multicolumn{3}{c}{500 K}\\ 
		\hline
		Mass [M$_\oplus$] & Radius$_\text{A}$ [R$_\oplus$] & Radius$_\text{Z}$ [R$_\oplus$] & $\delta_\text{R}$ [\%]& Radius$_\text{A}$ [R$_\oplus$] & Radius$_\text{Z}$ [R$_\oplus$] & $\delta_\text{R}$ [\%]\\
		\hline
		0.50 & 1.028 & 1.018 & -1.01 & 1.117 & 1.118 & 0.09 \\
		1.00 & 1.247 & 1.241 & -0.49 & 1.312 & 1.314 & 0.12 \\
		1.50 & 1.389 & 1.387 & -0.16 & 1.444 & 1.448 & 0.30 \\
		2.00 & 1.499 & 1.502 & 0.21 & 1.546 & 1.553 & 0.43 \\
		2.50 & 1.589 & 1.590 & 0.06 & 1.632 & 1.635 & 0.19 \\
		3.00 & 1.667 & 1.673 & 0.37 & 1.706 & 1.717 & 0.65 \\
		3.50 & 1.735 & 1.742 & 0.43 & 1.771 & 1.780 & 0.47 \\
		4.00 & 1.795 & 1.805 & 0.54 & 1.830 & 1.842 & 0.66 \\
		4.50 & 1.851 & 1.860 & 0.50 & 1.883 & 1.894 & 0.58 \\
		5.00 & 1.901 & 1.911 & 0.55 & 1.932 & 1.946 & 0.73 \\
		6.00 & 1.990 & 2.002 & 0.59 & 2.019 & 2.023 & 0.22 \\
		7.00 & 2.069 & 2.081 & 0.60 & 2.095 & 2.101 & 0.27 \\
		8.00 & 2.138 & 2.152 & 0.64 & 2.163 & 2.178 & 0.69 \\
		9.00 & 2.201 & 2.213 & 0.54 & 2.224 & 2.236 & 0.53 \\
		10.0 & 2.258 & 2.270 & 0.53 & 2.280 & 2.294 & 0.62 \\
		12.0 & 2.357 & 2.370 & 0.56 & 2.377 & 2.393 & 0.66 \\
		14.0 & 2.441 & 2.457 & 0.68 & 2.459 & 2.473 & 0.55 \\
		16.0 & 2.512 & 2.535 & 0.90 & 2.530 & 2.553 & 0.90 \\
		18.0 & 2.576 & 2.601 & 0.96 & 2.593 & 2.616 & 0.89 \\
		20.0 & 2.634 & 2.662 & 1.08 & 2.650 & 2.679 & 1.09 \\
		\hline
		T$_\text{surf}$&\multicolumn{3}{c}{700 K}  &\multicolumn{3}{c}{1000 K}\\ 
		\hline
		Mass [M$_\oplus$] & Radius$_\text{A}$ [R$_\oplus$] & Radius$_\text{Z}$ [R$_\oplus$] & $\delta_\text{R}$ [\%]& Radius$_\text{A}$ [R$_\oplus$] & Radius$_\text{Z}$ [R$_\oplus$] & $\delta_\text{R}$ [\%]\\
		\hline
	0.50 & 1.219 & 1.232 & 1.04 & 1.387 & 1.397 & 0.73 \\
	1.00 & 1.384 & 1.392 & 0.57 & 1.498 & 1.511 & 0.86 \\
	1.50 & 1.502 & 1.512 & 0.64 & 1.594 & 1.612 & 1.10 \\
	2.00 & 1.597 & 1.609 & 0.73 & 1.676 & 1.696 & 1.18 \\
	2.50 & 1.678 & 1.686 & 0.47 & 1.748 & 1.764 & 0.93 \\
	3.00 & 1.748 & 1.762 & 0.82 & 1.811 & 1.832 & 1.13 \\
	3.50 & 1.810 & 1.822 & 0.64 & 1.869 & 1.887 & 0.97 \\
	4.00 & 1.866 & 1.881 & 0.81 & 1.921 & 1.942 & 1.08 \\
	4.50 & 1.917 & 1.931 & 0.72 & 1.969 & 1.988 & 0.96 \\
	5.00 & 1.964 & 1.981 & 0.85 & 2.013 & 2.034 & 1.02 \\
	6.00 & 2.049 & 2.056 & 0.34 & 2.093 & 2.105 & 0.56 \\
	7.00 & 2.123 & 2.130 & 0.36 & 2.164 & 2.176 & 0.56 \\
	8.00 & 2.189 & 2.205 & 0.74 & 2.227 & 2.247 & 0.88 \\
	9.00 & 2.249 & 2.262 & 0.60 & 2.285 & 2.302 & 0.73 \\
	10.0 & 2.303 & 2.319 & 0.70 & 2.338 & 2.356 & 0.79 \\
	12.0 & 2.398 & 2.415 & 0.70 & 2.430 & 2.448 & 0.76 \\
	14.0 & 2.479 & 2.493 & 0.57 & 2.508 & 2.524 & 0.63 \\
	16.0 & 2.549 & 2.571 & 0.88 & 2.576 & 2.600 & 0.94 \\
	18.0 & 2.611 & 2.634 & 0.87 & 2.636 & 2.660 & 0.92 \\
	20.0 & 2.666 & 2.696 & 1.12 & 2.691 & 2.721 & 1.12 \\
		\hline
		
		\hline
\end{tabular}
\end{table*}
\begin{sidewaystable*}[h!]
	\caption{AQUA-equation of state for water, as a function of pressure and temperature. The complete table is available in electronic form at the CDS via anonymous ftp to cdsarc.u-strasbg.fr (130.79.128.5) or via \url{http://cdsweb.u-strasbg.fr/cgi-bin/qcat?J/A+A/}. It can also be downloaded from \url{https://github.com/mnijh/AQUA}. The columns are pressure P, temperature T, density $\rho$, adiabatic temperature gradient $\nabla_{Ad}$, specific entropy S, specific internal energy U, bulk speed of sound w, mean molecular weight $\mu$, ionisation fraction X$_\text{ion}$, dissociation fraction X$_\text{d}$ and the phase identifier. }
	\label{Tab:eos_tab1}
	\begin{tabular}{c c c c c c c c c c c}
		\hline\hline
		P [Pa] &T [K] & $\rho$ [kg/m$^3$] & $\nabla_\text{Ad}$ & S [J/(kg$\cdot$K)]&U [J/kg]& w [m/s] & $\mu$ [kg/mol] &  X$_\text{ion}$ &X$_\text{d}$ &PhaseID\\
		\hline 
		1.0E-01 & 1.00000000E+02 & 9.33039806E+02 & 3.72246653E-12 & 9.07515644E+02 & 4.22592576E+04 & 3.34777032E+03 & 1.80152680E-02 & 0.0 & 0.0 & -1 \\
		1.0E-01 & 1.02329299E+02 & 9.32971807E+02 & 3.87119922E-12 & 9.27848916E+02 & 4.43163494E+04 & 3.34627895E+03 & 1.80152680E-02 & 0.0 & 0.0 & -1 \\
		1.0E-01 & 1.04712855E+02 & 9.32898022E+02 & 4.02352149E-12 & 9.48595993E+02 & 4.64640048E+04 & 3.34467707E+03 & 1.80152680E-02 & 0.0 & 0.0 & -1 \\
		1.0E-01 & 1.07151931E+02 & 9.32818004E+02 & 4.17923014E-12 & 9.69761262E+02 & 4.87059756E+04 & 3.34295792E+03 & 1.80152680E-02 & 0.0 & 0.0 & -1 \\
		1.0E-01 & 1.09647820E+02 & 9.32731290E+02 & 4.33806557E-12 & 9.91349274E+02 & 5.10460816E+04 & 3.34111462E+03 & 1.80152680E-02 & 0.0 & 0.0 & -1 \\
		1.0E-01 & 1.12201845E+02 & 9.32637402E+02 & 4.49971313E-12 & 1.01336469E+03 & 5.34882049E+04 & 3.33914025E+03 & 1.80152680E-02 & 0.0 & 0.0 & -1 \\
		1.0E-01 & 1.14815362E+02 & 9.32535850E+02 & 4.66380534E-12 & 1.03581215E+03 & 5.60362767E+04 & 3.33702787E+03 & 1.80152680E-02 & 0.0 & 0.0 & -1 \\
		1.0E-01 & 1.17489755E+02 & 9.32426122E+02 & 4.82994164E-12 & 1.05869848E+03 & 5.86945310E+04 & 3.33477033E+03 & 1.80152680E-02 & 0.0 & 0.0 & -1 \\
		1.0E-01 & 1.20226443E+02 & 9.32307670E+02 & 4.99768651E-12 & 1.08202836E+03 & 6.14673847E+04 & 3.33236011E+03 & 1.80152680E-02 & 0.0 & 0.0 & -1 \\
		1.0E-01 & 1.23026877E+02 & 9.32179912E+02 & 5.16656914E-12 & 1.10580749E+03 & 6.43595786E+04 & 3.32978936E+03 & 1.80152680E-02 & 0.0 & 0.0 & -1 \\
		1.0E-01 & 1.25892541E+02 & 9.32042274E+02 & 5.33606781E-12 & 1.13004231E+03 & 6.73758848E+04 & 3.32705047E+03 & 1.80152680E-02 & 0.0 & 0.0 & -1 \\
		1.0E-01 & 1.28824955E+02 & 9.31894163E+02 & 5.50563196E-12 & 1.15473983E+03 & 7.05212749E+04 & 3.32413577E+03 & 1.80152680E-02 & 0.0 & 0.0 & -1 \\
		1.0E-01 & 1.31825674E+02 & 9.31734943E+02 & 5.67469371E-12 & 1.17990722E+03 & 7.38011901E+04 & 3.32103711E+03 & 1.80152680E-02 & 0.0 & 0.0 & -1 \\
		1.0E-01 & 1.34896288E+02 & 9.31563973E+02 & 5.84265974E-12 & 1.20555237E+03 & 7.72212545E+04 & 3.31774640E+03 & 1.80152680E-02 & 0.0 & 0.0 & -1 \\
		1.0E-01 & 1.38038426E+02 & 9.31380613E+02 & 6.00891853E-12 & 1.23168319E+03 & 8.07871764E+04 & 3.31425580E+03 & 1.80152680E-02 & 0.0 & 0.0 & -1 \\
		1.0E-01 & 1.41253754E+02 & 9.31184172E+02 & 6.17287121E-12 & 1.25831048E+03 & 8.45054606E+04 & 3.31055687E+03 & 1.80152680E-02 & 0.0 & 0.0 & -1 \\
		1.0E-01 & 1.44543977E+02 & 9.30973948E+02 & 6.33390229E-12 & 1.28544310E+03 & 8.83827349E+04 & 3.30664126E+03 & 1.80152680E-02 & 0.0 & 0.0 & -1 \\
		1.0E-01 & 1.47910839E+02 & 9.30749260E+02 & 6.49141542E-12 & 1.31309200E+03 & 9.24257674E+04 & 3.30250122E+03 & 1.80152680E-02 & 0.0 & 0.0 & -1 \\
		1.0E-01 & 1.51356125E+02 & 9.30509398E+02 & 6.64483403E-12 & 1.34126862E+03 & 9.66418650E+04 & 3.29812875E+03 & 1.80152680E-02 & 0.0 & 0.0 & -1 \\
		1.0E-01 & 1.54881662E+02 & 9.30253623E+02 & 6.79360346E-12 & 1.36998500E+03 & 1.01038863E+05 & 3.29351581E+03 & 1.80152680E-02 & 0.0 & 0.0 & -1 \\
		1.0E-01 & 1.58489319E+02 & 9.29981201E+02 & 6.93720585E-12 & 1.39925413E+03 & 1.05624899E+05 & 3.28865483E+03 & 1.80152680E-02 & 0.0 & 0.0 & -1 \\
		\hline
		\hline
	\end{tabular}
\end{sidewaystable*}

\begin{sidewaystable*}[h!]
	\caption{AQUA-equation of state for water, tabulated as a function of density and temperature. The complete table is available in electronic form at the CDS via anonymous ftp to cdsarc.u-strasbg.fr (130.79.128.5) or via \url{http://cdsweb.u-strasbg.fr/cgi-bin/qcat?J/A+A/}. It can also be downloaded from \url{https://github.com/mnijh/AQUA}. The columns are density $\rho$, temperature T, pressure P, adiabatic temperature gradient $\nabla_{Ad}$, specific entropy S, specific internal energy U, bulk speed of sound w, mean molecular weight $\mu$, ionisation fraction X$_\text{ion}$, dissociation fraction X$_\text{d}$ and the phase identifier. }
	\label{Tab:eos_tab2}
	\begin{tabular}{c c c c c c c c c c c}
		\hline\hline
		$\rho$ [kg/m$^3$] &T [K] &P [Pa] & $\nabla_\text{Ad}$ & S [J/(kg$\cdot$K)]&U [J/kg]& w [m/s] &$\mu$ [kg/mol] & X$_\text{ion}$ &X$_\text{d}$ &PhaseID\\
		\hline 
		1.0E-10 & 1.00000000E+02 & 1.0856626E-14 & 5.8730521E-10 & 2.8614436E+04 & 2.7667979E+06 & 2.4800860E+02  & 1.8015268E-02 & 0.0 & 0.0 &  0\\
		1.0E-10 & 1.02329299E+02 & 4.2621127E-14 & 2.2529884E-09 & 2.8026088E+04 & 2.7700289E+06 & 2.5087706E+02  & 1.8015268E-02 & 0.0 & 0.0 &  0\\
		1.0E-10 & 1.04712855E+02 & 1.6236629E-13 & 8.3868810E-09 & 2.7451601E+04 & 2.7733355E+06 & 2.5377924E+02  & 1.8015268E-02 & 0.0 & 0.0 &  0\\
		1.0E-10 & 1.07151931E+02 & 6.0061617E-13 & 3.0316360E-08 & 2.6890667E+04 & 2.7767194E+06 & 2.5671546E+02  & 1.8015268E-02 & 0.0 & 0.0 &  0\\
		1.0E-10 & 1.09647820E+02 & 2.1587929E-12 & 1.0648062E-07 & 2.6342988E+04 & 2.7801824E+06 & 2.5968607E+02  & 1.8015268E-02 & 0.0 & 0.0 &  0\\
		1.0E-10 & 1.12201845E+02 & 7.5442090E-12 & 3.6362709E-07 & 2.5808270E+04 & 2.7837262E+06 & 2.6269139E+02  & 1.8015268E-02 & 0.0 & 0.0 &  0\\
		1.0E-10 & 1.14815362E+02 & 2.5649249E-11 & 1.2081001E-06 & 2.5286229E+04 & 2.7873528E+06 & 2.6573178E+02  & 1.8015268E-02 & 0.0 & 0.0 &  0\\
		1.0E-10 & 1.17489755E+02 & 8.4890075E-11 & 3.9072668E-06 & 2.4776584E+04 & 2.7910640E+06 & 2.6880758E+02  & 1.8015268E-02 & 0.0 & 0.0 &  0\\
		1.0E-10 & 1.20226443E+02 & 2.7366385E-10 & 1.2309008E-05 & 2.4279064E+04 & 2.7948617E+06 & 2.7191915E+02  & 1.8015268E-02 & 0.0 & 0.0 &  0\\
		1.0E-10 & 1.23026877E+02 & 8.5981903E-10 & 3.7792286E-05 & 2.3793404E+04 & 2.7987481E+06 & 2.7506684E+02  & 1.8015268E-02 & 0.0 & 0.0 &  0\\
		1.0E-10 & 1.25892541E+02 & 2.6343331E-09 & 1.1315084E-04 & 2.3319342E+04 & 2.8027250E+06 & 2.7825101E+02  & 1.8015268E-02 & 0.0 & 0.0 &  0\\
		1.0E-10 & 1.28824955E+02 & 7.8749444E-09 & 3.3054112E-04 & 2.2856627E+04 & 2.8067948E+06 & 2.8147205E+02  & 1.8015268E-02 & 0.0 & 0.0 &  0\\
		1.0E-10 & 1.31825674E+02 & 2.2981105E-08 & 9.4262575E-04 & 2.2405011E+04 & 2.8109594E+06 & 2.8473031E+02  & 1.8015268E-02 & 0.0 & 0.0 &  0\\
		1.0E-10 & 1.34896288E+02 & 6.5504157E-08 & 2.6255899E-03 & 2.1964252E+04 & 2.8152212E+06 & 2.8802620E+02  & 1.8015268E-02 & 0.0 & 0.0 &  0\\
		1.0E-10 & 1.38038426E+02 & 1.8245839E-07 & 7.1467847E-03 & 2.1534115E+04 & 2.8195824E+06 & 2.9136016E+02  & 1.8015268E-02 & 0.0 & 0.0 &  0\\
		1.0E-10 & 1.41253754E+02 & 4.9690463E-07 & 1.9019885E-02 & 2.1114369E+04 & 2.8240453E+06 & 2.9473261E+02  & 1.8015268E-02 & 0.0 & 0.0 &  0\\
		1.0E-10 & 1.44543977E+02 & 1.3237576E-06 & 4.9514269E-02 & 2.0704790E+04 & 2.8286124E+06 & 2.9814400E+02  & 1.8015268E-02 & 0.0 & 0.0 &  0\\
		1.0E-10 & 1.47910839E+02 & 3.4512553E-06 & 1.2614936E-01 & 2.0305159E+04 & 2.8332861E+06 & 3.0159475E+02  & 1.8015268E-02 & 0.0 & 0.0 &  0\\
		1.0E-10 & 1.51356125E+02 & 6.9198149E-06 & 2.4716525E-01 & 2.0025076E+04 & 2.8380688E+06 & 3.0508532E+02  & 1.8015268E-02 & 0.0 & 0.0 &  3\\
		1.0E-10 & 1.54881662E+02 & 7.1480926E-06 & 2.4949790E-01 & 2.0052201E+04 & 2.8429632E+06 & 3.0861614E+02  & 1.8015268E-02 & 0.0 & 0.0 &  3\\
		1.0E-10 & 1.58489319E+02 & 7.3145943E-06 & 2.4948787E-01 & 2.0084168E+04 & 2.8479718E+06 & 3.1218767E+02  & 1.8015268E-02 & 0.0 & 0.0 &  3\\
		\hline
		\hline
	\end{tabular}
\end{sidewaystable*}	
\begin{sidewaystable*}[h!]
	\caption{AQUA-equation of state for water, tabulated as a function of density and internal energy. The complete table is available in electronic form at the CDS via anonymous ftp to cdsarc.u-strasbg.fr (130.79.128.5) or via \url{http://cdsweb.u-strasbg.fr/cgi-bin/qcat?J/A+A/}. It can also be downloaded from \url{https://github.com/mnijh/AQUA}. The columns are density $\rho$, specific internal energy U, pressure P, temperature T, adiabatic temperature gradient $\nabla_{Ad}$, specific entropy S,  bulk speed of sound w, mean molecular weight $\mu$, ionisation fraction X$_\text{ion}$, dissociation  fraction X$_\text{d}$ and the phase identifier.}
	\label{Tab:eos_tab3}
	\begin{tabular}{c c c c c c c c c c c}
		\hline\hline
		$\rho$ [kg/m$^3$]& U [J/kg] &P [Pa] &T [K] & $\nabla_\text{Ad}$ & S [J/(kg$\cdot$K)]& w [m/s] &$\mu$ [kg/mol] & X$_\text{ion}$ &X$_\text{d}$ &PhaseID\\
		\hline 
		1.0E-10 & 2.63412215E+06 & 0.00000000E+00 & 0.00000000E+00 & 0.00000000E+00 & 0.00000000E+00 & 0.00000000E+00 & 0.00000000E+00 & 0.0 & 0.0 & -99\\
		1.0E-10 & 2.69550653E+06 & 0.00000000E+00 & 0.00000000E+00 & 0.00000000E+00 & 0.00000000E+00 & 0.00000000E+00 & 0.00000000E+00 & 0.0 & 0.0 & -99 \\
		1.0E-10 & 2.75832139E+06 & 0.00000000E+00 & 0.00000000E+00 & 0.00000000E+00 & 0.00000000E+00 & 0.00000000E+00 & 0.00000000E+00 & 0.0 & 0.0 & -99\\
		1.0E-10 & 2.82260005E+06 & 3.69366399E-07 & 1.40210867E+02 & 1.42435924E-02 & 2.12485610E+04 & 2.93644733E+02 & 1.80152680E-02 & 0.0 & 0.0 & 0 \\
		1.0E-10 & 2.88837664E+06 & 8.65737390E-06 & 1.87581660E+02 & 2.49384210E-01 & 2.03182042E+04 & 3.39611093E+02 & 1.80152680E-02 & 0.0 & 0.0 & 3 \\
		1.0E-10 & 2.95568605E+06 & 1.08914713E-05 & 2.35988405E+02 & 2.48965205E-01 & 2.06373992E+04 & 3.80812181E+02 & 1.80152680E-02 & 0.0 & 0.0 & 3 \\
		1.0E-10 & 3.02456400E+06 & 1.31691745E-05 & 2.85343726E+02 & 2.47938351E-01 & 2.09024240E+04 & 4.18458151E+02 & 1.80152680E-02 & 0.0 & 0.0 & 3 \\
		1.0E-10 & 3.09504706E+06 & 1.54822082E-05 & 3.35462214E+02 & 2.46097352E-01 & 2.11299609E+04 & 4.53167346E+02 & 1.80152680E-02 & 0.0 & 0.0 & 3 \\
		1.0E-10 & 3.16717263E+06 & 1.78207107E-05 & 3.86128698E+02 & 2.43490948E-01 & 2.13301510E+04 & 4.85348812E+02 & 1.80152680E-02 & 0.0 & 0.0 & 3 \\
		1.0E-10 & 3.24097899E+06 & 2.01762418E-05 & 4.37169786E+02 & 2.40323965E-01 & 2.15096513E+04 & 5.15353726E+02 & 1.80152680E-02 & 0.0 & 0.0 & 3 \\
		1.0E-10 & 3.31650527E+06 & 2.25428557E-05 & 4.88442171E+02 & 2.36843925E-01 & 2.16729603E+04 & 5.43494715E+02 & 1.80152680E-02 & 0.0 & 0.0 & 3 \\
		1.0E-10 & 3.39379161E+06 & 2.49172634E-05 & 5.39891140E+02 & 2.33202730E-01 & 2.18233769E+04 & 5.70043531E+02 & 1.80152680E-02 & 0.0 & 0.0 & 3 \\
		1.0E-10 & 3.47287899E+06 & 2.72976002E-05 & 5.91468712E+02 & 2.29505095E-01 & 2.19632641E+04 & 5.95217911E+02 & 1.80152680E-02 & 0.0 & 0.0 & 3 \\
		1.0E-10 & 3.55380937E+06 & 2.96828490E-05 & 6.43146350E+02 & 2.25802764E-01 & 2.20944100E+04 & 6.19190962E+02 & 1.80152680E-02 & 0.0 & 0.0 & 3 \\
		1.0E-10 & 3.63662571E+06 & 3.20726732E-05 & 6.94932812E+02 & 2.22106379E-01 & 2.22182491E+04 & 6.42105321E+02 & 1.80152680E-02 & 0.0 & 0.0 & 3 \\
		1.0E-10 & 3.72137198E+06 & 3.44664991E-05 & 7.46795348E+02 & 2.18442912E-01 & 2.23358324E+04 & 6.64073259E+02 & 1.80152680E-02 & 0.0 & 0.0 & 3 \\
		1.0E-10 & 3.80809314E+06 & 3.68644353E-05 & 7.98755482E+02 & 2.14813460E-01 & 2.24480862E+04 & 6.85197474E+02 & 1.80152680E-02 & 0.0 & 0.0 & 3 \\
		1.0E-10 & 3.89683521E+06 & 3.92663320E-05 & 8.50796284E+02 & 2.11241093E-01 & 2.25556990E+04 & 7.05563513E+02 & 1.80152680E-02 & 0.0 & 0.0 & 3 \\
		1.0E-10 & 3.98764529E+06 & 4.16726579E-05 & 9.02940261E+02 & 2.07736011E-01 & 2.26592888E+04 & 7.25253038E+02 & 1.80152680E-02 & 0.0 & 0.0 & 3 \\
		1.0E-10 & 4.08057156E+06 & 4.40836721E-05 & 9.55176179E+02 & 2.04323114E-01 & 2.27593170E+04 & 7.44335624E+02 & 1.80152680E-02 & 0.0 & 0.0 & 3 \\
		\hline
		\hline
	\end{tabular}
\end{sidewaystable*}	
\end{document}